\documentclass[10pt]{article}

\usepackage[dvips,letterpaper]{geometry}

\usepackage{srcltx}

\usepackage{amsmath}
\usepackage{amssymb,amsfonts}

\usepackage{epsfig}
\usepackage{color}
\usepackage[usenames,dvipsnames]{xcolor}
\usepackage{subfigure}

\usepackage{booktabs}

\usepackage{setspace}
\usepackage{flushend}
\usepackage{multicol}

\usepackage{url}

\usepackage{enumerate}

\usepackage[short,12hr]{datetime}

\usepackage{hyperref}

\newtheorem{proposition}{Proposition}

\newtheorem{theorem}{Theorem}

\newtheorem{example}{Example}

\def\QED{\mbox{$\square$}}
\def\proof{\noindent{\it Proof:~}}
\def\endproof{\hspace*{\fill}~\QED\par\endtrivlist\unskip}

\newcommand{\down}{ {(\downarrow\!\!2)} }

\newcommand{\Z}{\stackrel{z}{\longleftrightarrow}}

\newcommand{\matlab}{{\sc Matlab}}

\title{%
Chebyshev Wavelets}

\title{On Filter Banks and Wavelets Based on \\Chebyshev Polynomials}

\author{%
R.~J.~Cintra%
\thanks{%
R. J. Cintra is with the
Signal Processing Group @ Stat,
Departamento de Estat\'istica, 
Universidade Federal de Pernambuco,
Brazil.
E-mail: \protect\url{rjdsc@dsp.ufpe.org}}
\quad
H.~M.~de~Oliveira%
\thanks{%
H. M. de Oliveira
is with the
Departamento de Eletr\^onica e Sistemas, 
Universidade Federal de Pernambuco,
Brazil.
E-mail: \protect\url{hmo@ufpe.br}}
\quad
L.~R.~Soares%
\thanks{%
L.~R.~Soares
is with 
Ag\^encia Nacional de Energia El\'etrica,
Brazil.}
}

\date{\today\ @ \currenttime}
\date{}

\begin{document}

\maketitle

\onehalfspacing

\doublespacing

\begin{abstract}
In this paper we introduce a new family of wavelets, named 
Chebyshev wavelets, 
which are derived from conventional
first and second kind 
Chebyshev polynomials.
Properties of Chebyshev filter banks are investigated, 
including orthogonality and perfect reconstruction conditions.
Chebyshev wavelets 
have compact support,
their filters possess good selectivity, but they are not orthogonal.
The convergence of the cascade algorithm of 
Chebyshev wavelets
is proved by using properties of Markov chains.
Computational implementation of these wavelets 
and some clear-cut applications are presented.
Proposed wavelets are suitable for signal denoising.
\end{abstract}

\begin{center}
{\small \textbf{Keywords}
\\
Discrete wavelets,
Chebyshev polynomials
}
\end{center}

\section{Introduction}

Sturm-Liouville theory encompasses a multitude of engineering and physics
problems~\cite{Arfk70}.
One particular and interesting case is that one related to Chebyshev 
differential equations.
Chebyshev polynomials of the first kind (Type I) of order $n$, $T_n(x)$,
satisfies the equation
$(1-x)\frac{\mathrm{d}^2y}{\mathrm{d}x^2} - x\frac{\mathrm{d}y}{\mathrm{d}x} + n^2 y = 0$
and
Chebyshev polynomials of second kind (Type II) of degree $n$, $U_n(x)$,
satisfies 
$(1-x^2)\frac{\mathrm{d}^2y}{\mathrm{d}x^2} - 3x\frac{\mathrm{d}y}{\mathrm{d}x}+n(n+2)y=0$.
Chebyshev polynomials form a complete set of orthogonal functions in the
interval $[-1,1]$
with weighting functions
$(1-x^2)^{-1/2}$ and
$(1-x^2)^{1/2}$, for the polynomials of first and second kind respectively.
Some special values are $T_n(1)=1$ and $T_{2n+1}=0$;
$U_n(1) = n+1$ and $U_{2n+1}(0)=0$.
Chebyshev polynomial also respect symmetry properties
$T_n(-x)=(-1)^n T_n(x)$ and $U_n(-x)=(-1)^n U_n(x)$~\cite{Arfk70,AbraSte68,KilgPres94}.

Chebyshev polynomials have many applications in numerical computations,
interpolation, series truncation and economization, to name a few.
In the past few years, connections between orthogonal polynomials and wavelet analysis 
have been explored, particularly a wavelet decomposition in $L^2(-1,1)$ has been
proposed~\cite{FiscPrest96,KilgPres94}.
Recently another approach has been investigated~\cite{Lira02,Lira03}:
the link between classical 
differential equation solutions---like 
Mathieu functions (elliptic cosine and sine) 
and 
Legendre polynomials---and wavelet design.

Exploring these connections, in this paper, 
we aim at 
developing new discrete wavelets
based on Chebyshev polynomials.
For such,
we consider the following
procedure:
(i)~smoothing filters based in Chebyshev polynomials are sought;
(ii)~a filter bank based on the obtained filters is derived;
(iii)~filter bank properties,
such as
perfect reconstruction and
orthogonality,
are examined;
and 
(iii)~the cascade iterative procedure is
applied on the proposed filter bank to create wavelets.

For the sake of notation, let us take 
the sequences $h[n]$ as the lowpass filters 
and
$g[n]$ as the highpass filters
(by convention $\sum_n h[n] = 1$ and $\sum_n g[n] = 0$).
The matrix $\mathbf{H}$ is the convolution matrix.
For the role of downsampling by two, it is adopted the operator $\down$.
Equality by definition is denoted by $\triangleq$.

\section{Chebyshev Filters}

In this section, we introduce
filter banks based on Chebyshev polynomials.
We examine the properties of such filters
for deriving new wavelets.

\subsection{First Kind Chebyshev Filters}

The well-known Chebyshev polynomials of 1st kind~$T_m(\cdot)$
are defined by a simple recurrence formulation~\cite{Arfk70}
furnished by:
\begin{equation*}
T_m(x) = 2x \cdot T_{m-1}(x) - T_{m-2}(x),
\end{equation*}
where
the initial conditions are
$T_0(x) \triangleq 1$ and $T_1(x) \triangleq x$.
Adopting
the variable change $x = \cos\omega$,
lowpass filters can be derived from these polynomials.
Indeed,
we have the new functions~\cite{AbraSte68}
\begin{equation}\label{cheb1:id}
T_m(\cos\omega) = \cos(m\omega),
\end{equation}
whose magnitude in the interval 
$[0,\pi]$ satisfies 
the
lowpass filter conditions for frequency response magnitude.
Polynomials can be considered 
as smoothing filters 
for wavelet generation through the cascade algorithm.

Smoothing filters $H(e^{j\omega})$ intended to be used for multiresolution analysis~\cite{Mall89} must
hold some specific conditions, such as $|H(e^{j0})| = 1$ and $|H(e^{j\pi})| = 0$.
In order to make Chebyshev polynomials useful for this kind of application,
a slight modification on $T_m(\cdot)$ is carried out so as to meet these constraints.
Taking only Chebyshev polynomials of odd order $m$, 
we can define the magnitude response of the smoothing filter as
\begin{equation*}
|H_m^{(1)}(e^{j\omega})| 
\triangleq 
|T_m(\cos(\omega/2))|
,
\quad 
\text{for odd $m$.}
\end{equation*}
Observe that these functions are naturally normalized.
Some examples can be seen in
Figure~\ref{plot:cheb:1}.

\begin{figure}[h]
  \begin{center}
     \begin{tabular}{cc}
       \hspace{-.95cm} \input{filtro-tn3.latex} & \hspace{-1.5cm} \input{filtro-tn5.latex}
     \end{tabular}
  \end{center}
  \vspace{-.5cm}
\caption{Plot of $|H_m^{(1)}(e^{j\omega})|$, for $m = 3, 5$, $\omega\in[0,\pi]$.}
\label{plot:cheb:1}
\end{figure}

In a previous work~\cite{Lira02}, wavelets based on Mathieu differential equations
were defined. %
The mathematical structure of Mathieu wavelets naturally induces a 
linear phase assignment $e^{-jm\omega}$ for the smoothing filter.
This approach was considered here.
After phase adjustment, 
we have the following
expression for the smoothing filter:
\begin{align*}
H_m^{(1)}(e^{j\omega}) \triangleq e^{-jm\omega/2} T_m(\cos(\omega/2)),\quad \text{$m$ odd.}
\end{align*}
Using Equation~\ref{cheb1:id},
we may easily recognize that
\begin{align*}
H_m^{(1)}(e^{j\omega}) =& e^{-jm\omega/2} T_m(\cos(\omega/2)) \\
=&e^{-jm\omega/2} \cos(m\omega/2) \\
=&\frac{1}{2} (1 + e^{-jm\omega}).
\end{align*}

Applying the inverse discrete-time Fourier transform
$H_m^{(1)}(e^{j\omega})$, we can find the discrete filter 
$h^{(1)}_m[n]$,
which is given by:
\begin{equation*}
h^{(1)}_m[n] = 
\begin{cases}
1/2, & n = 0, m, \\
0, &\text{otherwise.}
\end{cases}
\end{equation*}

\subsection{Second Kind Chebyshev Filters}

Now we examine another class of polynomials, namely the Chebyshev polynomials of 2nd kind.
This family of polynomials is also built from the same recurrence relation used
to derive the 1st kind ones. 
However, different initial conditions are set:
\begin{align*}
U_m(x) = 2x \cdot U_{m-1}(x) - U_{m-2}(x),
\end{align*}
for $U_0(x) = 1$ and $U_1(x)=2x$.
A variety of interesting properties and theorems on these polynomials 
can be found in~\cite{AbraSte68,Arfk70}.

Following similar steps and derivations as in the previous subsection,
we investigate the use of $U_m(x)$ in the definition 
of lowpass filters. 
This time, our aim is to construct new wavelets.

First, we adopt a usual variable change $x=\cos\omega$, yielding to~\cite[p.776]{AbraSte68}:
\begin{equation*}
U_m[\cos(\omega)] = \frac{\sin[(m+1)\omega]}{\sin(\omega)}.
\end{equation*}
Now we may consider the use of the modulus of these functions as
the magnitude response of lowpass filters.
However, one may not directly proceed in a such way, since
$|U_m(\cos\omega)|$ does not promptly satisfy 
lowpass filter conditions ($|H(e^{j0})|=1$ and $|H(e^{j\pi})|=0$).
To make this possible, a simple rule-of-thumb adjustment can be used.
Just as in the former 1st kind polynomial case, a scaling on 
the argument of $U_m(\cdot)$
by $1/2$ solves the problem, and makes $|H(e^{j\pi})|=0$.
The restriction of oddness for $m$ must be checked, otherwise the
proposed $\frac{1}{2}$-scaling on frequency cannot work.

In contrast with Chebyshev polynomials of 1st kind, 
the polynomials of 2nd kind are not naturally normalized.
The maximum value of $U_m(\cos\omega)$ is located at the peak of the main lobe
(vicinity of zero) and can be computed without effort:
\begin{equation*}
\lim_{\omega\to 0}U_m(\cos(\omega)) = 
\lim_{\omega\to 0} \frac{\sin[ (m+1)\omega]}{\sin(\omega)} = 
m+1.
\end{equation*}
Then a scaling factor of $\frac{1}{m+1}$ must be taken into consideration to normalize
the filter response. 
This adjustment redefines the magnitude of the frequency response to
\begin{equation*}
|H_m^{(2)}(e^{j\omega})| 
\triangleq 
\frac{1}{m+1}
\cdot
\left|
U_m
\left[
\cos\left(
\frac{\omega}{2}
\right)
\right]
\right|
,
\qquad \text{for odd $n$}.
\end{equation*}
This ensures that $|H_m^{(2)}(e^{j0})|=1$.
Illustrations of the frequency response magnitude of
$H_m^{(2)}(e^{j\omega})$ are shown in Figure~\ref{plot:cheb:2}.

\begin{figure}
  \begin{center}
     \begin{tabular}{cc}
       \hspace{-.95cm} \input{filtro-un5.latex} & \hspace{-1.5cm} \input{filtro-un7.latex}
     \end{tabular}
  \end{center}
  \vspace{-.5cm}
\caption{Plot of $|U_m(\cos(\omega/2))|$, for $m = 5, 7$, $\omega\in[0,\pi]$.}
\label{plot:cheb:2}
\end{figure}

The final, but crucial, step concerns phase assignment.
Again let us take a linear phase convenient choice~\cite{Lira02}.
Consequently,
the Chebyshev lowpass filters are completely specified by
\begin{equation*}
H_m^{(2)}(e^{j\omega})
\triangleq
 e^{-jm\omega/2}
\cdot
U_m
\left[
\cos\left(
\frac{\omega}{2}
\right)
\right]
.
\end{equation*}

Using now the fact that $U_m(\cos(\omega)) = \sin( (m+1)\omega)/\sin(\omega)$,
we can write the following:
\begin{equation*}
H_m^{(2)}(e^{j\omega})
=
\frac{1}{m+1}
e^{-jm\omega/2}
\frac{\sin( (m+1)\omega/2)}{\sin(\omega/2)}
.
\end{equation*}

This is the exact formulation of the
moving average filters or rectangular window~\cite{Oppe99}.
The impulse response~$h^{(2)}_m[n]$ of these filters are promptly derived:
\begin{equation}
\label{h:2}
h^{(2)}_m[n] = 
\begin{cases}
1/(m+1), & n = 0,\ldots, m, \\
0, &\text{otherwise.}
\end{cases}
\end{equation}

\section{Chebyshev Filter Banks}

\subsection{Type I Chebyshev Filter Banks}

We  use this filter $h^{(1)}_m[n]$ to define 
reconstruction and decomposition filter banks.
The relation among the highpass and lowpass 
filters of these two filter banks is 
well-esta\-blished~\cite{MisiMisi00,VettKova95,StraNguy96}
namely:
\begin{alignat}{2}\label{def:rec}
{h_r}^{(1)}[n] &= \sqrt{2} h^{(1)}_m[n],   &   \ \ 
{g_r}^{(1)}[n] &= \sqrt{2} (-1)^n h_r^{(1)}[m-n], \\\label{def:dec}
{h_d}^{(1)}[n] &= \sqrt{2} h_r^{(1)}[m-n], &    \ \ 
{g_d}^{(1)}[n] &= \sqrt{2} g_r^{(1)}[m-n],
\end{alignat}
for $n=0,\ldots,m$. 
Indexes $r$ and $d$ denote
reconstruction and decomposition filters,
respectively.

The filter banks based on lowpass filters $h^{(1)}_m[n]$ share perfect reconstruction property.
Let us use capital letters to denote $z$-transforms of time domain vectors.
Therefore $H_r^{(1)}$ is the $z$-transform of the lowpass reconstruction filter
$h_r^{(1)} \triangleq \sqrt{2} h^{(1)}_m$.
In a similar way, we may define the reconstruction and decomposition filter banks
$z$-transform by
$h_r^{(1)} \Z H_r^{(1)}$,
$g_r^{(1)} \Z G_r^{(1)}$,
$h_d^{(1)} \Z H_d^{(1)}$
and
$g_d^{(1)} \Z G_d^{(1)}$.

To achieve perfect reconstruction, a filter bank satisfies
alias cancellation and has no distortion properties. 
To ensure alias cancellation, we must have~\cite{Vett01} that:
\begin{equation*}
H_r^{(1)}(z)
H_d^{(1)}(-z)
+
G_r^{(1)}(z)
G_d^{(1)}(-z)
=
0.
\end{equation*}
Substituting these $z$-transforms by their corresponding explicit expressions and
taking into account that $m$ is odd, yields
\begin{equation*}
\label{alias:cancellation}
\begin{split}
\frac{1}{\sqrt{2}}(1+z^{-m})
&
\frac{1}{\sqrt{2}}(1+(-z)^{-m})
 + 
\frac{1}{\sqrt{2}}(-1+z^{-m})
\frac{1}{\sqrt{2}}(1-(-z)^{-m})
= \\
&\frac{1}{\sqrt{2}}(1+z^{-m})
\frac{1}{\sqrt{2}}(1-z^{-m})
 - \frac{1}{\sqrt{2}}(1-z^{-m})
\frac{1}{\sqrt{2}}(1+z^{-m})
= 0,
\end{split}
\end{equation*}
which asserts the alias cancellation property.
To ensure perfect reconstruction,
it is also required that
the filter banks introduce no distortion, i.e.,
only a delay is allowed~\cite{SmitBarn86}:
\begin{equation*}
\label{no:distortion}
H_r^{(1)}(z)
H_d^{(1)}(z)
+
G_r^{(1)}(z)
G_d^{(1)}(z)
=
2z^{-l}
.
\end{equation*}
After necessary manipulations,
we obtain:
\begin{equation*}
\begin{split}
\frac{1}{\sqrt{2}}(1+z^{-m})
\frac{1}{\sqrt{2}}(1+z^{-m})
 + 
\frac{1}{\sqrt{2}}(-1+z^{-m})
\frac{1}{\sqrt{2}}(1-z^{-m})
= 2 z^{-m}.
\end{split}
\end{equation*}
Notice that the filter bank delay is equal to $m$, 
exactly the order of the initially selected 
Chebyshev polynomial.

Another question to be examined is the orthogonality condition.
A filter bank is orthogonal if it satisfies even-shift
convolution ($\ast_2$)~\cite{SmitBarn86,StraNguy96}:
\begin{equation}\label{ortho}
h[n]\ast_2h[n] = \sum_k h[k] h[k-2n] = \delta[n],
\end{equation}
where $\delta[n]$ is the unit sample sequence.
It can be shown that the lowpass filter 
$h^{(1)}[n] = \frac{1}{2}\begin{bmatrix}1 & 0 & \cdots & 0 & 1\end{bmatrix}$ 
fulfills this orthogonality test.

Although these two desirable properties 
--- perfect reconstruction and orthogonality ---
are met, we will show that in a general manner the 
iterative process of the cascade algorithm using the filters
$h^{(1)}_m[n]$ does not lead to wavelets.
In other words, 
the limit of cascade algorithm is not a smooth function
and 
the algorithm does not converge in $L^2$.
The following theorem states 
a necessary and sufficient condition 
for iteration convergence~\cite{StraNguy96,Lawt91}.

\begin{theorem}[Smoothness]
\label{CondE}
Let $h[n]$ be a lowpass filter of length $m+1$ and $\mathbf{H}$ be its associated filter matrix.
If the infinite matrix $\mathbf{T} = \down 2\mathbf{H}\mathbf{H}^T$
has a centered submatrix $\mathbf{T}_{2m-1}$ of order $2m-1$,
such that 
all its eigenvalues satisfy $|\lambda|<1$ (except for a simple $\lambda = 1$), 
then the cascade algorithm converges in $L^2$ sense.
\end{theorem}

According to the definition given in Theorem~\ref{CondE}, by removing odd numbered
rows of $2\mathbf{H}\mathbf{H}^T$ 
(i.e., applying the decimation-by-2 operator $\down$), 
one can directly get $\mathbf{T}_{2m-1}$.
For Chebyshev polynomials of 1st kind, we have derived the filter
$\mathbf{h}^{(1)}_m = \frac{1}{2} \big[ 1\ \underbrace{0\ 0\ \cdots 0\  \ 0}_{\text{$m-1$ zeros.}}\ 1\big]$,
thus the rows of $2\mathbf{H}\mathbf{H}^T$ are
a stack of sequential single-shifted versions of the following vector:
\begin{equation*}
\begin{split}
\frac{1}{2} \begin{bmatrix}1 & 0 & \cdots & 0 & 1\end{bmatrix}
\ast
\begin{bmatrix}1 & 0 & \cdots & 0 & 1\end{bmatrix}
=
\frac{1}{2} \, \big[ 1\ \underbrace{0\ 0\ \cdots 0\  \ 0}_{\text{$m-1$ zeros.}}\ 2\ \underbrace{0\ 0\ \cdots 0\  \ 0}_{\text{$m-1$ zeros.}}\ 1 \big]
,
\end{split}
\end{equation*}
where $\ast$ denotes usual convolution.

Since the element~1 in this resulting vector is separated from the element~2 
by a even number of zeros $m-1$,
the odd-line elimination of $2\mathbf{H}\mathbf{H}^T$ will make
every column of $\mathbf{T}_{2m-1}$ have a single element~1 or
a pair of $1/2$, as it can be seen below:

\begin{equation*}
\mathbf{T}_{2m-1} = 
\frac{1}{2}
\left[
\begin{smallmatrix}
  &        &   &        &   &   &        &   &   &        &   &        &   \\ \medskip
0 & 1      & 0 & \cdots & 0 & 0 & 0      & 0 & 0 & \cdots & 0 & 0      & 0 \\ \medskip
  & \ddots &   & \ddots &   &   & \ddots &   &   & \ddots &   & \ddots &   \\ \medskip
0 & 0      & 0 & \cdots & 2 & 0 & 0      & 0 & 0 & \cdots & 0 & 1      & 0 \\ \medskip
0 & 0      & 0 & \cdots & 0 & 0 & 2      & 0 & 0 & \cdots & 0 & 0      & 0 \\ \medskip
0 & 1      & 0 & \cdots & 0 & 0 & 0      & 0 & 2 & \cdots & 0 & 0      & 0 \\ \medskip
  & \ddots &   & \ddots &   &   & \ddots &   &   & \ddots &   & \ddots &   \\ \medskip
0 & 0      & 0 & \cdots & 0 & 0 & 0      & 0 & 0 & \cdots & 0 & 1      & 0 \\ 
\end{smallmatrix}
\right].
\end{equation*}

By explicit 
computation of the eigenvalues,
the search for an $m$ which makes the matrix $\mathbf{T}_{2m-1}$ meet 
the conditions of Theorem~\ref{CondE} returned only one favorable case, for $m<256$.
This exception is $m=1$.
It is interesting to remark that when setting $m=1$,
the resulting 
$h^{(1)}_1[n] = \frac{1}{2}\begin{bmatrix}1 & 1\end{bmatrix}$ 
is the Haar filter bank, 
which makes the cascade algorithm generate the Haar wavelets.
Limited to our computational results, this is the only choice of Chebyshev 
polynomial that produces a wavelet.

\begin{example}
Let the lowpass filter 
$h^{(1)}_3[n] = \frac{1}{2}\begin{bmatrix}1 & 0 & 0& 1\end{bmatrix}$.
Since $m = 3$, the centered submatrix of 
$\down 2\cdot\mathbf{H}\cdot\mathbf{H}^T$ 
has order~$2m-1=5$.
Computing it, yields to
$$
\mathbf{T}_5
=
\frac{1}{2}
\begin{bmatrix}
0 & 1 & 0 & 0 & 0 \\
2 & 0 & 0 & 1 & 0 \\
0 & 0 & 2 & 0 & 0 \\
0 & 1 & 0 & 0 & 2 \\
0 & 0 & 0 & 1 & 0
\end{bmatrix},
$$
whose eigenvalues are $\pm\frac{1}{2}$ and $\pm1$ ($\lambda = 1$ has multiplicity of two).
We applied the cascade algorithm to this filter to visualize the emerging waveform
pattern (Figure~\ref{fig:cheb1:itera}).
\end{example}

\begin{figure}
  \begin{center}
     \begin{tabular}{cc}
       \hspace{-.75cm}
\setlength{\unitlength}{0.240900pt}
\ifx\plotpoint\undefined\newsavebox{\plotpoint}\fi
\begin{picture}(540,450)(0,0)
\font\gnuplot=cmr10 at 10pt
\gnuplot
\footnotesize
\sbox{\plotpoint}{\rule[-0.200pt]{0.400pt}{0.400pt}}%
\put(140.0,82.0){\rule[-0.200pt]{4.818pt}{0.400pt}}
\put(120,82){\makebox(0,0)[r]{-0.6}}
\put(459.0,82.0){\rule[-0.200pt]{4.818pt}{0.400pt}}
\put(140.0,164.0){\rule[-0.200pt]{4.818pt}{0.400pt}}
\put(120,164){\makebox(0,0)[r]{-0.3}}
\put(459.0,164.0){\rule[-0.200pt]{4.818pt}{0.400pt}}
\put(140.0,246.0){\rule[-0.200pt]{4.818pt}{0.400pt}}
\put(120,246){\makebox(0,0)[r]{0}}
\put(459.0,246.0){\rule[-0.200pt]{4.818pt}{0.400pt}}
\put(140.0,328.0){\rule[-0.200pt]{4.818pt}{0.400pt}}
\put(120,328){\makebox(0,0)[r]{0.3}}
\put(459.0,328.0){\rule[-0.200pt]{4.818pt}{0.400pt}}
\put(140.0,410.0){\rule[-0.200pt]{4.818pt}{0.400pt}}
\put(120,410){\makebox(0,0)[r]{0.6}}
\put(459.0,410.0){\rule[-0.200pt]{4.818pt}{0.400pt}}
\put(140.0,82.0){\rule[-0.200pt]{0.400pt}{4.818pt}}
\put(140,41){\makebox(0,0){0}}
\put(140.0,390.0){\rule[-0.200pt]{0.400pt}{4.818pt}}
\put(309.0,82.0){\rule[-0.200pt]{0.400pt}{4.818pt}}
\put(309,41){\makebox(0,0){5}}
\put(309.0,390.0){\rule[-0.200pt]{0.400pt}{4.818pt}}
\put(479.0,82.0){\rule[-0.200pt]{0.400pt}{4.818pt}}
\put(479,41){\makebox(0,0){10}}
\put(479.0,390.0){\rule[-0.200pt]{0.400pt}{4.818pt}}
\put(140.0,82.0){\rule[-0.200pt]{81.665pt}{0.400pt}}
\put(479.0,82.0){\rule[-0.200pt]{0.400pt}{79.015pt}}
\put(140.0,410.0){\rule[-0.200pt]{81.665pt}{0.400pt}}
\put(140.0,82.0){\rule[-0.200pt]{0.400pt}{79.015pt}}
\put(140,383){\usebox{\plotpoint}}
\multiput(140.58,375.89)(0.498,-2.030){65}{\rule{0.120pt}{1.712pt}}
\multiput(139.17,379.45)(34.000,-133.447){2}{\rule{0.400pt}{0.856pt}}
\multiput(208.58,246.00)(0.498,2.030){65}{\rule{0.120pt}{1.712pt}}
\multiput(207.17,246.00)(34.000,133.447){2}{\rule{0.400pt}{0.856pt}}
\multiput(242.58,375.89)(0.498,-2.030){65}{\rule{0.120pt}{1.712pt}}
\multiput(241.17,379.45)(34.000,-133.447){2}{\rule{0.400pt}{0.856pt}}
\put(174.0,246.0){\rule[-0.200pt]{8.191pt}{0.400pt}}
\multiput(309.58,238.89)(0.498,-2.030){65}{\rule{0.120pt}{1.712pt}}
\multiput(308.17,242.45)(34.000,-133.447){2}{\rule{0.400pt}{0.856pt}}
\multiput(343.58,109.00)(0.498,2.030){65}{\rule{0.120pt}{1.712pt}}
\multiput(342.17,109.00)(34.000,133.447){2}{\rule{0.400pt}{0.856pt}}
\put(276.0,246.0){\rule[-0.200pt]{7.950pt}{0.400pt}}
\multiput(411.58,238.89)(0.498,-2.030){65}{\rule{0.120pt}{1.712pt}}
\multiput(410.17,242.45)(34.000,-133.447){2}{\rule{0.400pt}{0.856pt}}
\put(377.0,246.0){\rule[-0.200pt]{8.191pt}{0.400pt}}
\end{picture} & \hspace{-1cm}
\setlength{\unitlength}{0.240900pt}
\ifx\plotpoint\undefined\newsavebox{\plotpoint}\fi
\begin{picture}(540,450)(0,0)
\font\gnuplot=cmr10 at 10pt
\gnuplot
\footnotesize
\sbox{\plotpoint}{\rule[-0.200pt]{0.400pt}{0.400pt}}%
\put(140.0,82.0){\rule[-0.200pt]{4.818pt}{0.400pt}}
\put(120,82){\makebox(0,0)[r]{-0.4}}
\put(459.0,82.0){\rule[-0.200pt]{4.818pt}{0.400pt}}
\put(140.0,164.0){\rule[-0.200pt]{4.818pt}{0.400pt}}
\put(120,164){\makebox(0,0)[r]{-0.2}}
\put(459.0,164.0){\rule[-0.200pt]{4.818pt}{0.400pt}}
\put(140.0,246.0){\rule[-0.200pt]{4.818pt}{0.400pt}}
\put(120,246){\makebox(0,0)[r]{0}}
\put(459.0,246.0){\rule[-0.200pt]{4.818pt}{0.400pt}}
\put(140.0,328.0){\rule[-0.200pt]{4.818pt}{0.400pt}}
\put(120,328){\makebox(0,0)[r]{0.2}}
\put(459.0,328.0){\rule[-0.200pt]{4.818pt}{0.400pt}}
\put(140.0,410.0){\rule[-0.200pt]{4.818pt}{0.400pt}}
\put(120,410){\makebox(0,0)[r]{0.4}}
\put(459.0,410.0){\rule[-0.200pt]{4.818pt}{0.400pt}}
\put(140.0,82.0){\rule[-0.200pt]{0.400pt}{4.818pt}}
\put(140,41){\makebox(0,0){0}}
\put(140.0,390.0){\rule[-0.200pt]{0.400pt}{4.818pt}}
\put(208.0,82.0){\rule[-0.200pt]{0.400pt}{4.818pt}}
\put(208,41){\makebox(0,0){5}}
\put(208.0,390.0){\rule[-0.200pt]{0.400pt}{4.818pt}}
\put(276.0,82.0){\rule[-0.200pt]{0.400pt}{4.818pt}}
\put(276,41){\makebox(0,0){10}}
\put(276.0,390.0){\rule[-0.200pt]{0.400pt}{4.818pt}}
\put(343.0,82.0){\rule[-0.200pt]{0.400pt}{4.818pt}}
\put(343,41){\makebox(0,0){15}}
\put(343.0,390.0){\rule[-0.200pt]{0.400pt}{4.818pt}}
\put(411.0,82.0){\rule[-0.200pt]{0.400pt}{4.818pt}}
\put(411,41){\makebox(0,0){20}}
\put(411.0,390.0){\rule[-0.200pt]{0.400pt}{4.818pt}}
\put(479.0,82.0){\rule[-0.200pt]{0.400pt}{4.818pt}}
\put(479,41){\makebox(0,0){25}}
\put(479.0,390.0){\rule[-0.200pt]{0.400pt}{4.818pt}}
\put(140.0,82.0){\rule[-0.200pt]{81.665pt}{0.400pt}}
\put(479.0,82.0){\rule[-0.200pt]{0.400pt}{79.015pt}}
\put(140.0,410.0){\rule[-0.200pt]{81.665pt}{0.400pt}}
\put(140.0,82.0){\rule[-0.200pt]{0.400pt}{79.015pt}}
\put(140,391){\usebox{\plotpoint}}
\multiput(140.58,373.39)(0.494,-5.308){25}{\rule{0.119pt}{4.243pt}}
\multiput(139.17,382.19)(14.000,-136.194){2}{\rule{0.400pt}{2.121pt}}
\multiput(167.58,246.00)(0.494,5.308){25}{\rule{0.119pt}{4.243pt}}
\multiput(166.17,246.00)(14.000,136.194){2}{\rule{0.400pt}{2.121pt}}
\multiput(181.58,372.06)(0.493,-5.730){23}{\rule{0.119pt}{4.562pt}}
\multiput(180.17,381.53)(13.000,-135.532){2}{\rule{0.400pt}{2.281pt}}
\put(154.0,246.0){\rule[-0.200pt]{3.132pt}{0.400pt}}
\multiput(208.58,246.00)(0.493,5.730){23}{\rule{0.119pt}{4.562pt}}
\multiput(207.17,246.00)(13.000,135.532){2}{\rule{0.400pt}{2.281pt}}
\multiput(221.58,373.39)(0.494,-5.308){25}{\rule{0.119pt}{4.243pt}}
\multiput(220.17,382.19)(14.000,-136.194){2}{\rule{0.400pt}{2.121pt}}
\put(194.0,246.0){\rule[-0.200pt]{3.373pt}{0.400pt}}
\multiput(248.58,246.00)(0.494,5.308){25}{\rule{0.119pt}{4.243pt}}
\multiput(247.17,246.00)(14.000,136.194){2}{\rule{0.400pt}{2.121pt}}
\multiput(262.58,373.39)(0.494,-5.308){25}{\rule{0.119pt}{4.243pt}}
\multiput(261.17,382.19)(14.000,-136.194){2}{\rule{0.400pt}{2.121pt}}
\put(235.0,246.0){\rule[-0.200pt]{3.132pt}{0.400pt}}
\multiput(289.58,228.39)(0.494,-5.308){25}{\rule{0.119pt}{4.243pt}}
\multiput(288.17,237.19)(14.000,-136.194){2}{\rule{0.400pt}{2.121pt}}
\multiput(303.58,101.00)(0.493,5.730){23}{\rule{0.119pt}{4.562pt}}
\multiput(302.17,101.00)(13.000,135.532){2}{\rule{0.400pt}{2.281pt}}
\put(276.0,246.0){\rule[-0.200pt]{3.132pt}{0.400pt}}
\multiput(330.58,227.06)(0.493,-5.730){23}{\rule{0.119pt}{4.562pt}}
\multiput(329.17,236.53)(13.000,-135.532){2}{\rule{0.400pt}{2.281pt}}
\multiput(343.58,101.00)(0.494,5.308){25}{\rule{0.119pt}{4.243pt}}
\multiput(342.17,101.00)(14.000,136.194){2}{\rule{0.400pt}{2.121pt}}
\put(316.0,246.0){\rule[-0.200pt]{3.373pt}{0.400pt}}
\multiput(371.58,227.06)(0.493,-5.730){23}{\rule{0.119pt}{4.562pt}}
\multiput(370.17,236.53)(13.000,-135.532){2}{\rule{0.400pt}{2.281pt}}
\multiput(384.58,101.00)(0.494,5.308){25}{\rule{0.119pt}{4.243pt}}
\multiput(383.17,101.00)(14.000,136.194){2}{\rule{0.400pt}{2.121pt}}
\put(357.0,246.0){\rule[-0.200pt]{3.373pt}{0.400pt}}
\multiput(411.58,228.39)(0.494,-5.308){25}{\rule{0.119pt}{4.243pt}}
\multiput(410.17,237.19)(14.000,-136.194){2}{\rule{0.400pt}{2.121pt}}
\put(398.0,246.0){\rule[-0.200pt]{3.132pt}{0.400pt}}
\end{picture} \\
       \hspace{-.75cm}
\setlength{\unitlength}{0.240900pt}
\ifx\plotpoint\undefined\newsavebox{\plotpoint}\fi
\begin{picture}(540,450)(0,0)
\font\gnuplot=cmr10 at 10pt
\gnuplot
\footnotesize
\sbox{\plotpoint}{\rule[-0.200pt]{0.400pt}{0.400pt}}%
\put(160.0,82.0){\rule[-0.200pt]{4.818pt}{0.400pt}}
\put(140,82){\makebox(0,0)[r]{-0.3}}
\put(459.0,82.0){\rule[-0.200pt]{4.818pt}{0.400pt}}
\put(160.0,164.0){\rule[-0.200pt]{4.818pt}{0.400pt}}
\put(140,164){\makebox(0,0)[r]{-0.15}}
\put(459.0,164.0){\rule[-0.200pt]{4.818pt}{0.400pt}}
\put(160.0,246.0){\rule[-0.200pt]{4.818pt}{0.400pt}}
\put(140,246){\makebox(0,0)[r]{0}}
\put(459.0,246.0){\rule[-0.200pt]{4.818pt}{0.400pt}}
\put(160.0,328.0){\rule[-0.200pt]{4.818pt}{0.400pt}}
\put(140,328){\makebox(0,0)[r]{0.15}}
\put(459.0,328.0){\rule[-0.200pt]{4.818pt}{0.400pt}}
\put(160.0,410.0){\rule[-0.200pt]{4.818pt}{0.400pt}}
\put(140,410){\makebox(0,0)[r]{0.3}}
\put(459.0,410.0){\rule[-0.200pt]{4.818pt}{0.400pt}}
\put(160.0,82.0){\rule[-0.200pt]{0.400pt}{4.818pt}}
\put(160,41){\makebox(0,0){0}}
\put(160.0,390.0){\rule[-0.200pt]{0.400pt}{4.818pt}}
\put(224.0,82.0){\rule[-0.200pt]{0.400pt}{4.818pt}}
\put(224,41){\makebox(0,0){10}}
\put(224.0,390.0){\rule[-0.200pt]{0.400pt}{4.818pt}}
\put(288.0,82.0){\rule[-0.200pt]{0.400pt}{4.818pt}}
\put(288,41){\makebox(0,0){20}}
\put(288.0,390.0){\rule[-0.200pt]{0.400pt}{4.818pt}}
\put(351.0,82.0){\rule[-0.200pt]{0.400pt}{4.818pt}}
\put(351,41){\makebox(0,0){30}}
\put(351.0,390.0){\rule[-0.200pt]{0.400pt}{4.818pt}}
\put(415.0,82.0){\rule[-0.200pt]{0.400pt}{4.818pt}}
\put(415,41){\makebox(0,0){40}}
\put(415.0,390.0){\rule[-0.200pt]{0.400pt}{4.818pt}}
\put(479.0,82.0){\rule[-0.200pt]{0.400pt}{4.818pt}}
\put(479,41){\makebox(0,0){50}}
\put(479.0,390.0){\rule[-0.200pt]{0.400pt}{4.818pt}}
\put(160.0,82.0){\rule[-0.200pt]{76.847pt}{0.400pt}}
\put(479.0,82.0){\rule[-0.200pt]{0.400pt}{79.015pt}}
\put(160.0,410.0){\rule[-0.200pt]{76.847pt}{0.400pt}}
\put(160.0,82.0){\rule[-0.200pt]{0.400pt}{79.015pt}}
\put(160,383){\usebox{\plotpoint}}
\multiput(160.59,344.67)(0.482,-12.334){9}{\rule{0.116pt}{9.233pt}}
\multiput(159.17,363.84)(6.000,-117.836){2}{\rule{0.400pt}{4.617pt}}
\multiput(173.59,246.00)(0.482,12.334){9}{\rule{0.116pt}{9.233pt}}
\multiput(172.17,246.00)(6.000,117.836){2}{\rule{0.400pt}{4.617pt}}
\multiput(179.59,350.09)(0.485,-10.409){11}{\rule{0.117pt}{7.929pt}}
\multiput(178.17,366.54)(7.000,-120.544){2}{\rule{0.400pt}{3.964pt}}
\put(166.0,246.0){\rule[-0.200pt]{1.686pt}{0.400pt}}
\multiput(192.59,246.00)(0.482,12.334){9}{\rule{0.116pt}{9.233pt}}
\multiput(191.17,246.00)(6.000,117.836){2}{\rule{0.400pt}{4.617pt}}
\multiput(198.59,350.09)(0.485,-10.409){11}{\rule{0.117pt}{7.929pt}}
\multiput(197.17,366.54)(7.000,-120.544){2}{\rule{0.400pt}{3.964pt}}
\put(186.0,246.0){\rule[-0.200pt]{1.445pt}{0.400pt}}
\multiput(211.59,246.00)(0.482,12.334){9}{\rule{0.116pt}{9.233pt}}
\multiput(210.17,246.00)(6.000,117.836){2}{\rule{0.400pt}{4.617pt}}
\multiput(217.59,350.09)(0.485,-10.409){11}{\rule{0.117pt}{7.929pt}}
\multiput(216.17,366.54)(7.000,-120.544){2}{\rule{0.400pt}{3.964pt}}
\put(205.0,246.0){\rule[-0.200pt]{1.445pt}{0.400pt}}
\multiput(230.59,246.00)(0.485,10.409){11}{\rule{0.117pt}{7.929pt}}
\multiput(229.17,246.00)(7.000,120.544){2}{\rule{0.400pt}{3.964pt}}
\multiput(237.59,344.67)(0.482,-12.334){9}{\rule{0.116pt}{9.233pt}}
\multiput(236.17,363.84)(6.000,-117.836){2}{\rule{0.400pt}{4.617pt}}
\put(224.0,246.0){\rule[-0.200pt]{1.445pt}{0.400pt}}
\multiput(249.59,246.00)(0.485,10.409){11}{\rule{0.117pt}{7.929pt}}
\multiput(248.17,246.00)(7.000,120.544){2}{\rule{0.400pt}{3.964pt}}
\multiput(256.59,344.67)(0.482,-12.334){9}{\rule{0.116pt}{9.233pt}}
\multiput(255.17,363.84)(6.000,-117.836){2}{\rule{0.400pt}{4.617pt}}
\put(243.0,246.0){\rule[-0.200pt]{1.445pt}{0.400pt}}
\multiput(268.59,246.00)(0.485,10.409){11}{\rule{0.117pt}{7.929pt}}
\multiput(267.17,246.00)(7.000,120.544){2}{\rule{0.400pt}{3.964pt}}
\multiput(275.59,344.67)(0.482,-12.334){9}{\rule{0.116pt}{9.233pt}}
\multiput(274.17,363.84)(6.000,-117.836){2}{\rule{0.400pt}{4.617pt}}
\put(262.0,246.0){\rule[-0.200pt]{1.445pt}{0.400pt}}
\multiput(288.59,246.00)(0.482,12.334){9}{\rule{0.116pt}{9.233pt}}
\multiput(287.17,246.00)(6.000,117.836){2}{\rule{0.400pt}{4.617pt}}
\multiput(294.59,344.67)(0.482,-12.334){9}{\rule{0.116pt}{9.233pt}}
\multiput(293.17,363.84)(6.000,-117.836){2}{\rule{0.400pt}{4.617pt}}
\put(281.0,246.0){\rule[-0.200pt]{1.686pt}{0.400pt}}
\multiput(307.59,207.67)(0.482,-12.334){9}{\rule{0.116pt}{9.233pt}}
\multiput(306.17,226.84)(6.000,-117.836){2}{\rule{0.400pt}{4.617pt}}
\multiput(313.59,109.00)(0.482,12.334){9}{\rule{0.116pt}{9.233pt}}
\multiput(312.17,109.00)(6.000,117.836){2}{\rule{0.400pt}{4.617pt}}
\put(300.0,246.0){\rule[-0.200pt]{1.686pt}{0.400pt}}
\multiput(326.59,207.67)(0.482,-12.334){9}{\rule{0.116pt}{9.233pt}}
\multiput(325.17,226.84)(6.000,-117.836){2}{\rule{0.400pt}{4.617pt}}
\multiput(332.59,109.00)(0.485,10.409){11}{\rule{0.117pt}{7.929pt}}
\multiput(331.17,109.00)(7.000,120.544){2}{\rule{0.400pt}{3.964pt}}
\put(319.0,246.0){\rule[-0.200pt]{1.686pt}{0.400pt}}
\multiput(345.59,207.67)(0.482,-12.334){9}{\rule{0.116pt}{9.233pt}}
\multiput(344.17,226.84)(6.000,-117.836){2}{\rule{0.400pt}{4.617pt}}
\multiput(351.59,109.00)(0.485,10.409){11}{\rule{0.117pt}{7.929pt}}
\multiput(350.17,109.00)(7.000,120.544){2}{\rule{0.400pt}{3.964pt}}
\put(339.0,246.0){\rule[-0.200pt]{1.445pt}{0.400pt}}
\multiput(364.59,213.09)(0.485,-10.409){11}{\rule{0.117pt}{7.929pt}}
\multiput(363.17,229.54)(7.000,-120.544){2}{\rule{0.400pt}{3.964pt}}
\multiput(371.59,109.00)(0.482,12.334){9}{\rule{0.116pt}{9.233pt}}
\multiput(370.17,109.00)(6.000,117.836){2}{\rule{0.400pt}{4.617pt}}
\put(358.0,246.0){\rule[-0.200pt]{1.445pt}{0.400pt}}
\multiput(383.59,213.09)(0.485,-10.409){11}{\rule{0.117pt}{7.929pt}}
\multiput(382.17,229.54)(7.000,-120.544){2}{\rule{0.400pt}{3.964pt}}
\multiput(390.59,109.00)(0.482,12.334){9}{\rule{0.116pt}{9.233pt}}
\multiput(389.17,109.00)(6.000,117.836){2}{\rule{0.400pt}{4.617pt}}
\put(377.0,246.0){\rule[-0.200pt]{1.445pt}{0.400pt}}
\multiput(402.59,213.09)(0.485,-10.409){11}{\rule{0.117pt}{7.929pt}}
\multiput(401.17,229.54)(7.000,-120.544){2}{\rule{0.400pt}{3.964pt}}
\multiput(409.59,109.00)(0.482,12.334){9}{\rule{0.116pt}{9.233pt}}
\multiput(408.17,109.00)(6.000,117.836){2}{\rule{0.400pt}{4.617pt}}
\put(396.0,246.0){\rule[-0.200pt]{1.445pt}{0.400pt}}
\multiput(422.59,207.67)(0.482,-12.334){9}{\rule{0.116pt}{9.233pt}}
\multiput(421.17,226.84)(6.000,-117.836){2}{\rule{0.400pt}{4.617pt}}
\multiput(428.59,109.00)(0.482,12.334){9}{\rule{0.116pt}{9.233pt}}
\multiput(427.17,109.00)(6.000,117.836){2}{\rule{0.400pt}{4.617pt}}
\put(415.0,246.0){\rule[-0.200pt]{1.686pt}{0.400pt}}
\multiput(441.59,207.67)(0.482,-12.334){9}{\rule{0.116pt}{9.233pt}}
\multiput(440.17,226.84)(6.000,-117.836){2}{\rule{0.400pt}{4.617pt}}
\put(434.0,246.0){\rule[-0.200pt]{1.686pt}{0.400pt}}
\end{picture} & \hspace{-1cm}\input{cheb134.latex} \\
     \end{tabular}
  \end{center}
\caption{Waveform pattern emerged from first order Chebyshev filters iteration 
for $m = 3$ and 1, 2, 3 and 4 iterations.
There is no convergence in $L^2$.}
\label{fig:cheb1:itera}
\end{figure}

\subsection{Type II Chebyshev Filter Banks}

Taking Equation~\ref{h:2} as a starting point, 
we are now in a position to carry on some investigation
on Type II Chebyshev filter banks.

Based on $h^{(2)}_m[n]$ and using similar definitions for 
the reconstruction and decomposition filters 
as done before (cf.~\eqref{def:rec} and~\eqref{def:dec}), 
we may find the following $z$-transforms for
$h_r^{(2)}[n]$,
$g_r^{(2)}[n]$,
$h_d^{(2)}[n]$,
and
$g_d^{(2)}[n]$:
\begin{align*}
H_r^{(2)}(z) =& \frac{\sqrt 2}{m+1} \sum_{i=0}^m z^{-i}, 
\qquad
G_r^{(2)}(z) = \frac{\sqrt 2}{m+1} \sum_{i=0}^m (-1)^i z^{-1}, \\
H_d^{(2)}(z) =& \frac{\sqrt 2}{m+1} \sum_{i=0}^m z^{-i}, 
\qquad
G_d^{(2)}(z) = \frac{\sqrt 2}{m+1} \sum_{i=0}^m -(-1)^{i} z^{-1}.
\end{align*}

Let us begin examining perfect reconstruction questions.
As stated before, a filter satisfying both alias cancellation and
no distortion satisfies 
the following conditions:
\begin{align}
H_r^{(2)}(z)
H_d^{(2)}(-z)
+
G_r^{(2)}(z)
G_d^{(2)}(-z)
&=
0
, 
\nonumber
\\
\label{no:distortion:2}
H_r^{(2)}(z)
H_d^{(2)}(z)
+
G_r^{(2)}(z)
G_d^{(2)}(z)
&=
2z^{-l},
\end{align}
respectively.
After some routine algebraic
manipulation, 
we find that alias cancellation property 
is satisfied,
as shown below:
\begin{equation*}
\begin{split}
\frac{\sqrt 2}{m+1} \sum_{i=0}^m z^{-i}
\frac{\sqrt 2}{m+1} \sum_{i=0}^m (-z)^{-i}
+ 
 \frac{\sqrt 2}{m+1} \sum_{i=0}^m (-1)^i z^{-i}
\frac{\sqrt 2}{m+1} \sum_{i=0}^m -(-1)^{i} (-z)^{-i}
= 
\frac{2}{(m+1)^2} 
\\
\times
\left(
 \sum_{i=0}^m z^{-i}
 \sum_{i=0}^m (-1)^i z^{-i}
\right. 
- 
\left.
 \sum_{i=0}^m (-z)^{-i}
 \sum_{i=0}^m z^{-i}
\right)
= 
0
. 
\end{split}
\end{equation*}
However, 
after an application of Equation~\ref{no:distortion:2}, we find that
\begin{equation*}
\begin{split}
H_r^{(2)}(z)
H_d^{(2)}(z)
&+
G_r^{(2)}(z)
G_d^{(2)}(z)
= 
\left(
\frac{1-z^{-(m+1)}}{2}
\right)^2
\frac{z}{\left( 1+z^{-2} \right)^2}.
\end{split}
\end{equation*}
Since this is not in the form $2z^{-l}$, 
we conclude that
such a filter bank introduces some
distortion. 

It is easy to see that $h^{(2)}[n]$ does not verify Equation~\ref{ortho}, therefore
there is no orthogonality.
It remains to examine whether this filter bank class produces a convergent
smoothing (regular) wave or not. 

\begin{proposition}
\label{proposition-ood-converges}
Filter banks based on odd order Chebyshev polynomial of 2nd kind 
satisfy Theorem~\ref{CondE}.
\end{proposition}
\proof
In the appendix, we supply a proof
for
following proposition.
\endproof

Figure~\ref{itera:cheb:2} displays some results derived by
the iterative cascade algorithm, depicting the formation of
a wavelet function with compact support.

\begin{figure}
\centering
     \begin{tabular}{cc}
       \hspace{-.75cm} 
\setlength{\unitlength}{0.240900pt}
\ifx\plotpoint\undefined\newsavebox{\plotpoint}\fi
\begin{picture}(540,450)(0,0)
\font\gnuplot=cmr10 at 10pt
\gnuplot
\footnotesize
\sbox{\plotpoint}{\rule[-0.200pt]{0.400pt}{0.400pt}}%
\put(160.0,82.0){\rule[-0.200pt]{4.818pt}{0.400pt}}
\put(140,82){\makebox(0,0)[r]{-0.03}}
\put(459.0,82.0){\rule[-0.200pt]{4.818pt}{0.400pt}}
\put(160.0,137.0){\rule[-0.200pt]{4.818pt}{0.400pt}}
\put(140,137){\makebox(0,0)[r]{-0.02}}
\put(459.0,137.0){\rule[-0.200pt]{4.818pt}{0.400pt}}
\put(160.0,191.0){\rule[-0.200pt]{4.818pt}{0.400pt}}
\put(140,191){\makebox(0,0)[r]{-0.01}}
\put(459.0,191.0){\rule[-0.200pt]{4.818pt}{0.400pt}}
\put(160.0,246.0){\rule[-0.200pt]{4.818pt}{0.400pt}}
\put(140,246){\makebox(0,0)[r]{0}}
\put(459.0,246.0){\rule[-0.200pt]{4.818pt}{0.400pt}}
\put(160.0,301.0){\rule[-0.200pt]{4.818pt}{0.400pt}}
\put(140,301){\makebox(0,0)[r]{0.01}}
\put(459.0,301.0){\rule[-0.200pt]{4.818pt}{0.400pt}}
\put(160.0,355.0){\rule[-0.200pt]{4.818pt}{0.400pt}}
\put(140,355){\makebox(0,0)[r]{0.02}}
\put(459.0,355.0){\rule[-0.200pt]{4.818pt}{0.400pt}}
\put(160.0,410.0){\rule[-0.200pt]{4.818pt}{0.400pt}}
\put(140,410){\makebox(0,0)[r]{0.03}}
\put(459.0,410.0){\rule[-0.200pt]{4.818pt}{0.400pt}}
\put(160.0,82.0){\rule[-0.200pt]{0.400pt}{4.818pt}}
\put(160,41){\makebox(0,0){0}}
\put(160.0,390.0){\rule[-0.200pt]{0.400pt}{4.818pt}}
\put(240.0,82.0){\rule[-0.200pt]{0.400pt}{4.818pt}}
\put(240,41){\makebox(0,0){9}}
\put(240.0,390.0){\rule[-0.200pt]{0.400pt}{4.818pt}}
\put(319.0,82.0){\rule[-0.200pt]{0.400pt}{4.818pt}}
\put(319,41){\makebox(0,0){18}}
\put(319.0,390.0){\rule[-0.200pt]{0.400pt}{4.818pt}}
\put(399.0,82.0){\rule[-0.200pt]{0.400pt}{4.818pt}}
\put(399,41){\makebox(0,0){27}}
\put(399.0,390.0){\rule[-0.200pt]{0.400pt}{4.818pt}}
\put(479.0,82.0){\rule[-0.200pt]{0.400pt}{4.818pt}}
\put(479,41){\makebox(0,0){36}}
\put(479.0,390.0){\rule[-0.200pt]{0.400pt}{4.818pt}}
\put(160.0,82.0){\rule[-0.200pt]{76.847pt}{0.400pt}}
\put(479.0,82.0){\rule[-0.200pt]{0.400pt}{79.015pt}}
\put(160.0,410.0){\rule[-0.200pt]{76.847pt}{0.400pt}}
\put(160.0,82.0){\rule[-0.200pt]{0.400pt}{79.015pt}}
\put(160,318){\usebox{\plotpoint}}
\multiput(169.59,318.00)(0.489,4.106){15}{\rule{0.118pt}{3.256pt}}
\multiput(168.17,318.00)(9.000,64.243){2}{\rule{0.400pt}{1.628pt}}
\put(160.0,318.0){\rule[-0.200pt]{2.168pt}{0.400pt}}
\multiput(204.59,375.49)(0.489,-4.106){15}{\rule{0.118pt}{3.256pt}}
\multiput(203.17,382.24)(9.000,-64.243){2}{\rule{0.400pt}{1.628pt}}
\put(178.0,389.0){\rule[-0.200pt]{6.263pt}{0.400pt}}
\multiput(240.59,318.00)(0.489,4.106){15}{\rule{0.118pt}{3.256pt}}
\multiput(239.17,318.00)(9.000,64.243){2}{\rule{0.400pt}{1.628pt}}
\put(213.0,318.0){\rule[-0.200pt]{6.504pt}{0.400pt}}
\multiput(257.59,375.49)(0.489,-4.106){15}{\rule{0.118pt}{3.256pt}}
\multiput(256.17,382.24)(9.000,-64.243){2}{\rule{0.400pt}{1.628pt}}
\put(249.0,389.0){\rule[-0.200pt]{1.927pt}{0.400pt}}
\multiput(275.59,291.02)(0.489,-8.358){15}{\rule{0.118pt}{6.500pt}}
\multiput(274.17,304.51)(9.000,-130.509){2}{\rule{0.400pt}{3.250pt}}
\put(266.0,318.0){\rule[-0.200pt]{2.168pt}{0.400pt}}
\multiput(311.59,174.00)(0.488,9.475){13}{\rule{0.117pt}{7.300pt}}
\multiput(310.17,174.00)(8.000,128.848){2}{\rule{0.400pt}{3.650pt}}
\put(284.0,174.0){\rule[-0.200pt]{6.504pt}{0.400pt}}
\multiput(346.59,291.02)(0.489,-8.358){15}{\rule{0.118pt}{6.500pt}}
\multiput(345.17,304.51)(9.000,-130.509){2}{\rule{0.400pt}{3.250pt}}
\put(319.0,318.0){\rule[-0.200pt]{6.504pt}{0.400pt}}
\multiput(364.59,160.49)(0.489,-4.106){15}{\rule{0.118pt}{3.256pt}}
\multiput(363.17,167.24)(9.000,-64.243){2}{\rule{0.400pt}{1.628pt}}
\put(355.0,174.0){\rule[-0.200pt]{2.168pt}{0.400pt}}
\multiput(382.59,103.00)(0.488,4.654){13}{\rule{0.117pt}{3.650pt}}
\multiput(381.17,103.00)(8.000,63.424){2}{\rule{0.400pt}{1.825pt}}
\put(373.0,103.0){\rule[-0.200pt]{2.168pt}{0.400pt}}
\multiput(417.59,160.49)(0.489,-4.106){15}{\rule{0.118pt}{3.256pt}}
\multiput(416.17,167.24)(9.000,-64.243){2}{\rule{0.400pt}{1.628pt}}
\put(390.0,174.0){\rule[-0.200pt]{6.504pt}{0.400pt}}
\multiput(452.59,103.00)(0.489,4.106){15}{\rule{0.118pt}{3.256pt}}
\multiput(451.17,103.00)(9.000,64.243){2}{\rule{0.400pt}{1.628pt}}
\put(426.0,103.0){\rule[-0.200pt]{6.263pt}{0.400pt}}
\put(461.0,174.0){\rule[-0.200pt]{2.168pt}{0.400pt}}
\end{picture} & \hspace{-1cm}
\setlength{\unitlength}{0.240900pt}
\ifx\plotpoint\undefined\newsavebox{\plotpoint}\fi
\begin{picture}(540,450)(0,0)
\font\gnuplot=cmr10 at 10pt
\gnuplot
\footnotesize
\sbox{\plotpoint}{\rule[-0.200pt]{0.400pt}{0.400pt}}%
\put(180.0,82.0){\rule[-0.200pt]{4.818pt}{0.400pt}}
\put(160,82){\makebox(0,0)[r]{-0.015}}
\put(459.0,82.0){\rule[-0.200pt]{4.818pt}{0.400pt}}
\put(180.0,137.0){\rule[-0.200pt]{4.818pt}{0.400pt}}
\put(160,137){\makebox(0,0)[r]{-0.01}}
\put(459.0,137.0){\rule[-0.200pt]{4.818pt}{0.400pt}}
\put(180.0,191.0){\rule[-0.200pt]{4.818pt}{0.400pt}}
\put(160,191){\makebox(0,0)[r]{-0.005}}
\put(459.0,191.0){\rule[-0.200pt]{4.818pt}{0.400pt}}
\put(180.0,246.0){\rule[-0.200pt]{4.818pt}{0.400pt}}
\put(160,246){\makebox(0,0)[r]{0}}
\put(459.0,246.0){\rule[-0.200pt]{4.818pt}{0.400pt}}
\put(180.0,301.0){\rule[-0.200pt]{4.818pt}{0.400pt}}
\put(160,301){\makebox(0,0)[r]{0.005}}
\put(459.0,301.0){\rule[-0.200pt]{4.818pt}{0.400pt}}
\put(180.0,355.0){\rule[-0.200pt]{4.818pt}{0.400pt}}
\put(160,355){\makebox(0,0)[r]{0.01}}
\put(459.0,355.0){\rule[-0.200pt]{4.818pt}{0.400pt}}
\put(180.0,410.0){\rule[-0.200pt]{4.818pt}{0.400pt}}
\put(160,410){\makebox(0,0)[r]{0.015}}
\put(459.0,410.0){\rule[-0.200pt]{4.818pt}{0.400pt}}
\put(180.0,82.0){\rule[-0.200pt]{0.400pt}{4.818pt}}
\put(180,41){\makebox(0,0){0}}
\put(180.0,390.0){\rule[-0.200pt]{0.400pt}{4.818pt}}
\put(330.0,82.0){\rule[-0.200pt]{0.400pt}{4.818pt}}
\put(330,41){\makebox(0,0){25}}
\put(330.0,390.0){\rule[-0.200pt]{0.400pt}{4.818pt}}
\put(479.0,82.0){\rule[-0.200pt]{0.400pt}{4.818pt}}
\put(479,41){\makebox(0,0){50}}
\put(479.0,390.0){\rule[-0.200pt]{0.400pt}{4.818pt}}
\put(180.0,82.0){\rule[-0.200pt]{72.029pt}{0.400pt}}
\put(479.0,82.0){\rule[-0.200pt]{0.400pt}{79.015pt}}
\put(180.0,410.0){\rule[-0.200pt]{72.029pt}{0.400pt}}
\put(180.0,82.0){\rule[-0.200pt]{0.400pt}{79.015pt}}
\put(180,306){\usebox{\plotpoint}}
\multiput(186.59,306.00)(0.482,5.463){9}{\rule{0.116pt}{4.167pt}}
\multiput(185.17,306.00)(6.000,52.352){2}{\rule{0.400pt}{2.083pt}}
\put(180.0,306.0){\rule[-0.200pt]{1.445pt}{0.400pt}}
\multiput(270.59,349.70)(0.482,-5.463){9}{\rule{0.116pt}{4.167pt}}
\multiput(269.17,358.35)(6.000,-52.352){2}{\rule{0.400pt}{2.083pt}}
\put(192.0,367.0){\rule[-0.200pt]{18.790pt}{0.400pt}}
\multiput(282.59,288.98)(0.482,-5.373){9}{\rule{0.116pt}{4.100pt}}
\multiput(281.17,297.49)(6.000,-51.490){2}{\rule{0.400pt}{2.050pt}}
\put(276.0,306.0){\rule[-0.200pt]{1.445pt}{0.400pt}}
\multiput(365.59,228.98)(0.482,-5.373){9}{\rule{0.116pt}{4.100pt}}
\multiput(364.17,237.49)(6.000,-51.490){2}{\rule{0.400pt}{2.050pt}}
\put(288.0,246.0){\rule[-0.200pt]{18.549pt}{0.400pt}}
\multiput(377.59,168.70)(0.482,-5.463){9}{\rule{0.116pt}{4.167pt}}
\multiput(376.17,177.35)(6.000,-52.352){2}{\rule{0.400pt}{2.083pt}}
\put(371.0,186.0){\rule[-0.200pt]{1.445pt}{0.400pt}}
\multiput(461.59,125.00)(0.482,5.463){9}{\rule{0.116pt}{4.167pt}}
\multiput(460.17,125.00)(6.000,52.352){2}{\rule{0.400pt}{2.083pt}}
\put(383.0,125.0){\rule[-0.200pt]{18.790pt}{0.400pt}}
\put(467.0,186.0){\rule[-0.200pt]{1.445pt}{0.400pt}}
\end{picture} \\
       \hspace{-.75cm} 
\setlength{\unitlength}{0.240900pt}
\ifx\plotpoint\undefined\newsavebox{\plotpoint}\fi
\begin{picture}(540,450)(0,0)
\font\gnuplot=cmr10 at 10pt
\gnuplot
\footnotesize
\sbox{\plotpoint}{\rule[-0.200pt]{0.400pt}{0.400pt}}%
\put(160.0,82.0){\rule[-0.200pt]{4.818pt}{0.400pt}}
\put(140,82){\makebox(0,0)[r]{-0.02}}
\put(459.0,82.0){\rule[-0.200pt]{4.818pt}{0.400pt}}
\put(160.0,164.0){\rule[-0.200pt]{4.818pt}{0.400pt}}
\put(140,164){\makebox(0,0)[r]{-0.01}}
\put(459.0,164.0){\rule[-0.200pt]{4.818pt}{0.400pt}}
\put(160.0,246.0){\rule[-0.200pt]{4.818pt}{0.400pt}}
\put(140,246){\makebox(0,0)[r]{0}}
\put(459.0,246.0){\rule[-0.200pt]{4.818pt}{0.400pt}}
\put(160.0,328.0){\rule[-0.200pt]{4.818pt}{0.400pt}}
\put(140,328){\makebox(0,0)[r]{0.01}}
\put(459.0,328.0){\rule[-0.200pt]{4.818pt}{0.400pt}}
\put(160.0,410.0){\rule[-0.200pt]{4.818pt}{0.400pt}}
\put(140,410){\makebox(0,0)[r]{0.02}}
\put(459.0,410.0){\rule[-0.200pt]{4.818pt}{0.400pt}}
\put(160.0,82.0){\rule[-0.200pt]{0.400pt}{4.818pt}}
\put(160,41){\makebox(0,0){0}}
\put(160.0,390.0){\rule[-0.200pt]{0.400pt}{4.818pt}}
\put(240.0,82.0){\rule[-0.200pt]{0.400pt}{4.818pt}}
\put(240,41){\makebox(0,0){20}}
\put(240.0,390.0){\rule[-0.200pt]{0.400pt}{4.818pt}}
\put(319.0,82.0){\rule[-0.200pt]{0.400pt}{4.818pt}}
\put(319,41){\makebox(0,0){40}}
\put(319.0,390.0){\rule[-0.200pt]{0.400pt}{4.818pt}}
\put(399.0,82.0){\rule[-0.200pt]{0.400pt}{4.818pt}}
\put(399,41){\makebox(0,0){60}}
\put(399.0,390.0){\rule[-0.200pt]{0.400pt}{4.818pt}}
\put(479.0,82.0){\rule[-0.200pt]{0.400pt}{4.818pt}}
\put(479,41){\makebox(0,0){80}}
\put(479.0,390.0){\rule[-0.200pt]{0.400pt}{4.818pt}}
\put(160.0,82.0){\rule[-0.200pt]{76.847pt}{0.400pt}}
\put(479.0,82.0){\rule[-0.200pt]{0.400pt}{79.015pt}}
\put(160.0,410.0){\rule[-0.200pt]{76.847pt}{0.400pt}}
\put(160.0,82.0){\rule[-0.200pt]{0.400pt}{79.015pt}}
\put(160,271){\usebox{\plotpoint}}
\multiput(164.60,271.00)(0.468,3.698){5}{\rule{0.113pt}{2.700pt}}
\multiput(163.17,271.00)(4.000,20.396){2}{\rule{0.400pt}{1.350pt}}
\put(160.0,271.0){\rule[-0.200pt]{0.964pt}{0.400pt}}
\multiput(172.60,297.00)(0.468,7.207){5}{\rule{0.113pt}{5.100pt}}
\multiput(171.17,297.00)(4.000,39.415){2}{\rule{0.400pt}{2.550pt}}
\put(168.0,297.0){\rule[-0.200pt]{0.964pt}{0.400pt}}
\multiput(180.60,347.00)(0.468,3.698){5}{\rule{0.113pt}{2.700pt}}
\multiput(179.17,347.00)(4.000,20.396){2}{\rule{0.400pt}{1.350pt}}
\put(176.0,347.0){\rule[-0.200pt]{0.964pt}{0.400pt}}
\multiput(188.60,373.00)(0.468,3.552){5}{\rule{0.113pt}{2.600pt}}
\multiput(187.17,373.00)(4.000,19.604){2}{\rule{0.400pt}{1.300pt}}
\put(184.0,373.0){\rule[-0.200pt]{0.964pt}{0.400pt}}
\multiput(204.60,387.21)(0.468,-3.552){5}{\rule{0.113pt}{2.600pt}}
\multiput(203.17,392.60)(4.000,-19.604){2}{\rule{0.400pt}{1.300pt}}
\put(192.0,398.0){\rule[-0.200pt]{2.891pt}{0.400pt}}
\multiput(212.60,361.79)(0.468,-3.698){5}{\rule{0.113pt}{2.700pt}}
\multiput(211.17,367.40)(4.000,-20.396){2}{\rule{0.400pt}{1.350pt}}
\put(208.0,373.0){\rule[-0.200pt]{0.964pt}{0.400pt}}
\multiput(220.60,336.21)(0.468,-3.552){5}{\rule{0.113pt}{2.600pt}}
\multiput(219.17,341.60)(4.000,-19.604){2}{\rule{0.400pt}{1.300pt}}
\put(216.0,347.0){\rule[-0.200pt]{0.964pt}{0.400pt}}
\multiput(236.60,322.00)(0.468,3.552){5}{\rule{0.113pt}{2.600pt}}
\multiput(235.17,322.00)(4.000,19.604){2}{\rule{0.400pt}{1.300pt}}
\put(224.0,322.0){\rule[-0.200pt]{2.891pt}{0.400pt}}
\multiput(244.60,347.00)(0.468,3.698){5}{\rule{0.113pt}{2.700pt}}
\multiput(243.17,347.00)(4.000,20.396){2}{\rule{0.400pt}{1.350pt}}
\put(240.0,347.0){\rule[-0.200pt]{0.964pt}{0.400pt}}
\multiput(260.60,361.79)(0.468,-3.698){5}{\rule{0.113pt}{2.700pt}}
\multiput(259.17,367.40)(4.000,-20.396){2}{\rule{0.400pt}{1.350pt}}
\put(248.0,373.0){\rule[-0.200pt]{2.891pt}{0.400pt}}
\multiput(268.60,315.04)(0.468,-11.009){5}{\rule{0.113pt}{7.700pt}}
\multiput(267.17,331.02)(4.000,-60.018){2}{\rule{0.400pt}{3.850pt}}
\put(264.0,347.0){\rule[-0.200pt]{0.964pt}{0.400pt}}
\multiput(276.60,249.83)(0.468,-7.207){5}{\rule{0.113pt}{5.100pt}}
\multiput(275.17,260.41)(4.000,-39.415){2}{\rule{0.400pt}{2.550pt}}
\put(272.0,271.0){\rule[-0.200pt]{0.964pt}{0.400pt}}
\multiput(284.60,199.41)(0.468,-7.354){5}{\rule{0.113pt}{5.200pt}}
\multiput(283.17,210.21)(4.000,-40.207){2}{\rule{0.400pt}{2.600pt}}
\put(280.0,221.0){\rule[-0.200pt]{0.964pt}{0.400pt}}
\multiput(300.60,170.00)(0.468,7.354){5}{\rule{0.113pt}{5.200pt}}
\multiput(299.17,170.00)(4.000,40.207){2}{\rule{0.400pt}{2.600pt}}
\put(288.0,170.0){\rule[-0.200pt]{2.891pt}{0.400pt}}
\multiput(308.60,221.00)(0.468,7.207){5}{\rule{0.113pt}{5.100pt}}
\multiput(307.17,221.00)(4.000,39.415){2}{\rule{0.400pt}{2.550pt}}
\put(304.0,221.0){\rule[-0.200pt]{0.964pt}{0.400pt}}
\multiput(316.61,271.00)(0.447,11.179){3}{\rule{0.108pt}{6.900pt}}
\multiput(315.17,271.00)(3.000,36.679){2}{\rule{0.400pt}{3.450pt}}
\put(312.0,271.0){\rule[-0.200pt]{0.964pt}{0.400pt}}
\multiput(331.60,300.41)(0.468,-7.354){5}{\rule{0.113pt}{5.200pt}}
\multiput(330.17,311.21)(4.000,-40.207){2}{\rule{0.400pt}{2.600pt}}
\put(319.0,322.0){\rule[-0.200pt]{2.891pt}{0.400pt}}
\multiput(339.60,249.83)(0.468,-7.207){5}{\rule{0.113pt}{5.100pt}}
\multiput(338.17,260.41)(4.000,-39.415){2}{\rule{0.400pt}{2.550pt}}
\put(335.0,271.0){\rule[-0.200pt]{0.964pt}{0.400pt}}
\multiput(347.60,189.04)(0.468,-11.009){5}{\rule{0.113pt}{7.700pt}}
\multiput(346.17,205.02)(4.000,-60.018){2}{\rule{0.400pt}{3.850pt}}
\put(343.0,221.0){\rule[-0.200pt]{0.964pt}{0.400pt}}
\multiput(355.60,133.79)(0.468,-3.698){5}{\rule{0.113pt}{2.700pt}}
\multiput(354.17,139.40)(4.000,-20.396){2}{\rule{0.400pt}{1.350pt}}
\put(351.0,145.0){\rule[-0.200pt]{0.964pt}{0.400pt}}
\multiput(371.60,119.00)(0.468,3.698){5}{\rule{0.113pt}{2.700pt}}
\multiput(370.17,119.00)(4.000,20.396){2}{\rule{0.400pt}{1.350pt}}
\put(359.0,119.0){\rule[-0.200pt]{2.891pt}{0.400pt}}
\multiput(379.60,145.00)(0.468,3.552){5}{\rule{0.113pt}{2.600pt}}
\multiput(378.17,145.00)(4.000,19.604){2}{\rule{0.400pt}{1.300pt}}
\put(375.0,145.0){\rule[-0.200pt]{0.964pt}{0.400pt}}
\multiput(395.60,159.21)(0.468,-3.552){5}{\rule{0.113pt}{2.600pt}}
\multiput(394.17,164.60)(4.000,-19.604){2}{\rule{0.400pt}{1.300pt}}
\put(383.0,170.0){\rule[-0.200pt]{2.891pt}{0.400pt}}
\multiput(403.60,133.79)(0.468,-3.698){5}{\rule{0.113pt}{2.700pt}}
\multiput(402.17,139.40)(4.000,-20.396){2}{\rule{0.400pt}{1.350pt}}
\put(399.0,145.0){\rule[-0.200pt]{0.964pt}{0.400pt}}
\multiput(411.60,108.21)(0.468,-3.552){5}{\rule{0.113pt}{2.600pt}}
\multiput(410.17,113.60)(4.000,-19.604){2}{\rule{0.400pt}{1.300pt}}
\put(407.0,119.0){\rule[-0.200pt]{0.964pt}{0.400pt}}
\multiput(427.60,94.00)(0.468,3.552){5}{\rule{0.113pt}{2.600pt}}
\multiput(426.17,94.00)(4.000,19.604){2}{\rule{0.400pt}{1.300pt}}
\put(415.0,94.0){\rule[-0.200pt]{2.891pt}{0.400pt}}
\multiput(435.60,119.00)(0.468,3.698){5}{\rule{0.113pt}{2.700pt}}
\multiput(434.17,119.00)(4.000,20.396){2}{\rule{0.400pt}{1.350pt}}
\put(431.0,119.0){\rule[-0.200pt]{0.964pt}{0.400pt}}
\multiput(443.60,145.00)(0.468,7.207){5}{\rule{0.113pt}{5.100pt}}
\multiput(442.17,145.00)(4.000,39.415){2}{\rule{0.400pt}{2.550pt}}
\put(439.0,145.0){\rule[-0.200pt]{0.964pt}{0.400pt}}
\multiput(451.60,195.00)(0.468,3.698){5}{\rule{0.113pt}{2.700pt}}
\multiput(450.17,195.00)(4.000,20.396){2}{\rule{0.400pt}{1.350pt}}
\put(447.0,195.0){\rule[-0.200pt]{0.964pt}{0.400pt}}
\put(455.0,221.0){\rule[-0.200pt]{0.964pt}{0.400pt}}
\end{picture} & \hspace{-1cm}
\setlength{\unitlength}{0.240900pt}
\ifx\plotpoint\undefined\newsavebox{\plotpoint}\fi
\begin{picture}(540,450)(0,0)
\font\gnuplot=cmr10 at 10pt
\gnuplot
\footnotesize
\sbox{\plotpoint}{\rule[-0.200pt]{0.400pt}{0.400pt}}%
\put(180.0,82.0){\rule[-0.200pt]{4.818pt}{0.400pt}}
\put(160,82){\makebox(0,0)[r]{-0.008}}
\put(459.0,82.0){\rule[-0.200pt]{4.818pt}{0.400pt}}
\put(180.0,164.0){\rule[-0.200pt]{4.818pt}{0.400pt}}
\put(160,164){\makebox(0,0)[r]{-0.004}}
\put(459.0,164.0){\rule[-0.200pt]{4.818pt}{0.400pt}}
\put(180.0,246.0){\rule[-0.200pt]{4.818pt}{0.400pt}}
\put(160,246){\makebox(0,0)[r]{0}}
\put(459.0,246.0){\rule[-0.200pt]{4.818pt}{0.400pt}}
\put(180.0,328.0){\rule[-0.200pt]{4.818pt}{0.400pt}}
\put(160,328){\makebox(0,0)[r]{0.004}}
\put(459.0,328.0){\rule[-0.200pt]{4.818pt}{0.400pt}}
\put(180.0,410.0){\rule[-0.200pt]{4.818pt}{0.400pt}}
\put(160,410){\makebox(0,0)[r]{0.008}}
\put(459.0,410.0){\rule[-0.200pt]{4.818pt}{0.400pt}}
\put(180.0,82.0){\rule[-0.200pt]{0.400pt}{4.818pt}}
\put(180,41){\makebox(0,0){0}}
\put(180.0,390.0){\rule[-0.200pt]{0.400pt}{4.818pt}}
\put(255.0,82.0){\rule[-0.200pt]{0.400pt}{4.818pt}}
\put(255,41){\makebox(0,0){30}}
\put(255.0,390.0){\rule[-0.200pt]{0.400pt}{4.818pt}}
\put(330.0,82.0){\rule[-0.200pt]{0.400pt}{4.818pt}}
\put(330,41){\makebox(0,0){60}}
\put(330.0,390.0){\rule[-0.200pt]{0.400pt}{4.818pt}}
\put(404.0,82.0){\rule[-0.200pt]{0.400pt}{4.818pt}}
\put(404,41){\makebox(0,0){90}}
\put(404.0,390.0){\rule[-0.200pt]{0.400pt}{4.818pt}}
\put(479.0,82.0){\rule[-0.200pt]{0.400pt}{4.818pt}}
\put(479,41){\makebox(0,0){120}}
\put(479.0,390.0){\rule[-0.200pt]{0.400pt}{4.818pt}}
\put(180.0,82.0){\rule[-0.200pt]{72.029pt}{0.400pt}}
\put(479.0,82.0){\rule[-0.200pt]{0.400pt}{79.015pt}}
\put(180.0,410.0){\rule[-0.200pt]{72.029pt}{0.400pt}}
\put(180.0,82.0){\rule[-0.200pt]{0.400pt}{79.015pt}}
\put(180,266){\usebox{\plotpoint}}
\multiput(182.61,266.00)(0.447,4.258){3}{\rule{0.108pt}{2.767pt}}
\multiput(181.17,266.00)(3.000,14.258){2}{\rule{0.400pt}{1.383pt}}
\put(180.0,266.0){\rule[-0.200pt]{0.482pt}{0.400pt}}
\multiput(187.61,286.00)(0.447,8.723){3}{\rule{0.108pt}{5.433pt}}
\multiput(186.17,286.00)(3.000,28.723){2}{\rule{0.400pt}{2.717pt}}
\put(185.0,286.0){\rule[-0.200pt]{0.482pt}{0.400pt}}
\multiput(192.61,326.00)(0.447,8.723){3}{\rule{0.108pt}{5.433pt}}
\multiput(191.17,326.00)(3.000,28.723){2}{\rule{0.400pt}{2.717pt}}
\put(190.0,326.0){\rule[-0.200pt]{0.482pt}{0.400pt}}
\multiput(197.61,366.00)(0.447,4.258){3}{\rule{0.108pt}{2.767pt}}
\multiput(196.17,366.00)(3.000,14.258){2}{\rule{0.400pt}{1.383pt}}
\put(195.0,366.0){\rule[-0.200pt]{0.482pt}{0.400pt}}
\multiput(202.61,386.00)(0.447,4.258){3}{\rule{0.108pt}{2.767pt}}
\multiput(201.17,386.00)(3.000,14.258){2}{\rule{0.400pt}{1.383pt}}
\put(200.0,386.0){\rule[-0.200pt]{0.482pt}{0.400pt}}
\multiput(257.61,394.52)(0.447,-4.258){3}{\rule{0.108pt}{2.767pt}}
\multiput(256.17,400.26)(3.000,-14.258){2}{\rule{0.400pt}{1.383pt}}
\put(205.0,406.0){\rule[-0.200pt]{12.527pt}{0.400pt}}
\multiput(262.61,374.52)(0.447,-4.258){3}{\rule{0.108pt}{2.767pt}}
\multiput(261.17,380.26)(3.000,-14.258){2}{\rule{0.400pt}{1.383pt}}
\put(260.0,386.0){\rule[-0.200pt]{0.482pt}{0.400pt}}
\multiput(267.61,343.45)(0.447,-8.723){3}{\rule{0.108pt}{5.433pt}}
\multiput(266.17,354.72)(3.000,-28.723){2}{\rule{0.400pt}{2.717pt}}
\put(265.0,366.0){\rule[-0.200pt]{0.482pt}{0.400pt}}
\multiput(272.61,303.45)(0.447,-8.723){3}{\rule{0.108pt}{5.433pt}}
\multiput(271.17,314.72)(3.000,-28.723){2}{\rule{0.400pt}{2.717pt}}
\put(270.0,326.0){\rule[-0.200pt]{0.482pt}{0.400pt}}
\multiput(277.61,274.52)(0.447,-4.258){3}{\rule{0.108pt}{2.767pt}}
\multiput(276.17,280.26)(3.000,-14.258){2}{\rule{0.400pt}{1.383pt}}
\put(275.0,286.0){\rule[-0.200pt]{0.482pt}{0.400pt}}
\multiput(282.61,254.52)(0.447,-4.258){3}{\rule{0.108pt}{2.767pt}}
\multiput(281.17,260.26)(3.000,-14.258){2}{\rule{0.400pt}{1.383pt}}
\put(280.0,266.0){\rule[-0.200pt]{0.482pt}{0.400pt}}
\put(337.17,226){\rule{0.400pt}{4.100pt}}
\multiput(336.17,237.49)(2.000,-11.490){2}{\rule{0.400pt}{2.050pt}}
\put(285.0,246.0){\rule[-0.200pt]{12.527pt}{0.400pt}}
\put(342.17,206){\rule{0.400pt}{4.100pt}}
\multiput(341.17,217.49)(2.000,-11.490){2}{\rule{0.400pt}{2.050pt}}
\put(339.0,226.0){\rule[-0.200pt]{0.723pt}{0.400pt}}
\put(347.17,166){\rule{0.400pt}{8.100pt}}
\multiput(346.17,189.19)(2.000,-23.188){2}{\rule{0.400pt}{4.050pt}}
\put(344.0,206.0){\rule[-0.200pt]{0.723pt}{0.400pt}}
\put(352.17,126){\rule{0.400pt}{8.100pt}}
\multiput(351.17,149.19)(2.000,-23.188){2}{\rule{0.400pt}{4.050pt}}
\put(349.0,166.0){\rule[-0.200pt]{0.723pt}{0.400pt}}
\put(357.17,106){\rule{0.400pt}{4.100pt}}
\multiput(356.17,117.49)(2.000,-11.490){2}{\rule{0.400pt}{2.050pt}}
\put(354.0,126.0){\rule[-0.200pt]{0.723pt}{0.400pt}}
\put(362.17,86){\rule{0.400pt}{4.100pt}}
\multiput(361.17,97.49)(2.000,-11.490){2}{\rule{0.400pt}{2.050pt}}
\put(359.0,106.0){\rule[-0.200pt]{0.723pt}{0.400pt}}
\put(417.17,86){\rule{0.400pt}{4.100pt}}
\multiput(416.17,86.00)(2.000,11.490){2}{\rule{0.400pt}{2.050pt}}
\put(364.0,86.0){\rule[-0.200pt]{12.768pt}{0.400pt}}
\put(422.17,106){\rule{0.400pt}{4.100pt}}
\multiput(421.17,106.00)(2.000,11.490){2}{\rule{0.400pt}{2.050pt}}
\put(419.0,106.0){\rule[-0.200pt]{0.723pt}{0.400pt}}
\put(427.17,126){\rule{0.400pt}{8.100pt}}
\multiput(426.17,126.00)(2.000,23.188){2}{\rule{0.400pt}{4.050pt}}
\put(424.0,126.0){\rule[-0.200pt]{0.723pt}{0.400pt}}
\put(432.17,166){\rule{0.400pt}{8.100pt}}
\multiput(431.17,166.00)(2.000,23.188){2}{\rule{0.400pt}{4.050pt}}
\put(429.0,166.0){\rule[-0.200pt]{0.723pt}{0.400pt}}
\put(437.17,206){\rule{0.400pt}{4.100pt}}
\multiput(436.17,206.00)(2.000,11.490){2}{\rule{0.400pt}{2.050pt}}
\put(434.0,206.0){\rule[-0.200pt]{0.723pt}{0.400pt}}
\put(439.0,226.0){\rule[-0.200pt]{0.723pt}{0.400pt}}
\end{picture} \\
       \hspace{-.75cm} \input{cheb256.latex} & \hspace{-1cm}\input{cheb276.latex} \\
       (a) & (b)
     \end{tabular}
\caption{Second order Chebyshev wavelets in 2, 3 and 4 iterations, (a) for $m = 5$ and (b) for $m=7$.}
\label{itera:cheb:2}
\end{figure}

\begin{example}
Take the Chebyshev 2nd kind filter of order 3, 
$h^{(2)}_3 = \frac{1}{4}\begin{bmatrix}1 & 1 & 1 & 1\end{bmatrix}$.
Constructing the centered submatrix of $\mathbf{T} = \down 2\mathbf{H}\mathbf{H}^T$,
we have:
$$
\mathbf{T}_5
=
\frac{1}{8}
\begin{bmatrix}
2 & 1 & 0 & 0 & 0 \\
4 & 3 & 2 & 1 & 0 \\
2 & 3 & 4 & 3 & 2 \\
0 & 1 & 2 & 3 & 4 \\
0 & 0 & 0 & 1 & 2
\end{bmatrix}.
$$
Since all eigenvalues --- $1$, $\frac{1}{2}$, $\frac{1}{4}$ and 0 (double) ---
are less than one (except one), the regularity is assured.
\end{example}

\subsection{Filter Selectivity}

We can tune the selectivity of the $H^{(2)}$ filter by a judicious scaling adjustment.
Instead of taking the one-half scaling ($\omega/2$) on the second kind Chebyshev polynomial, we
could examine a more general modification.
Let the $H'^{(2)}(\omega)$ be the generalized second kind Chebyshev polynomial
smoothing filter defined by
\begin{align*}
H'^{(2)}_{m,r}(e^{j\omega})
=
e^{-j m \omega \left(   \frac{r}{m+1} \right)  }
U_m\left( \cos \left( \frac{r}{m+1}\omega \right)  \right),
\end{align*}
where $r = (2k+1)\frac{m+1}{2}$, $k = 0, 1, 2, \ldots$.

Observe that the previous smoothing filter $H^{(2)}(e^{j\omega})$ discussed in the
previous subsection is a special case of this new filter.
This can be checked by taking $r = \frac{m+1}{2}$.

\begin{figure}
\centering
     \begin{tabular}{c}
        \hspace{-.75cm}\input{cheb2-dec.latex}  \\
     \end{tabular}
\caption{Selectivity adjustment vs. change of Chebyshev polynomial.
The solid thin line represents $|H_{5,9}'^{(2)}(\omega)|$ and
the bold line is the plot of $|H_{5}^{(2)}(\omega)|$.
Upsampling factor of 3 ($\uparrow\!3$).}
\label{diferente}
\end{figure}

\begin{equation*}
h'^{(2)}[n] = 
\begin{cases}
h^{(2)}
\left[
\frac{n}
{ \left(\frac{2r}{m+1}\right) }
\right], & \text{if $\frac{2r}{m+1}|n$}, \\
0, &\text{otherwise.}
\end{cases}
\end{equation*}

\noindent
Figure~\ref{diferente} contains a elucidative example of this.

\section{Application and Discussion} %

\begin{figure}
\centering
\subfigure[]{
\epsfig{file=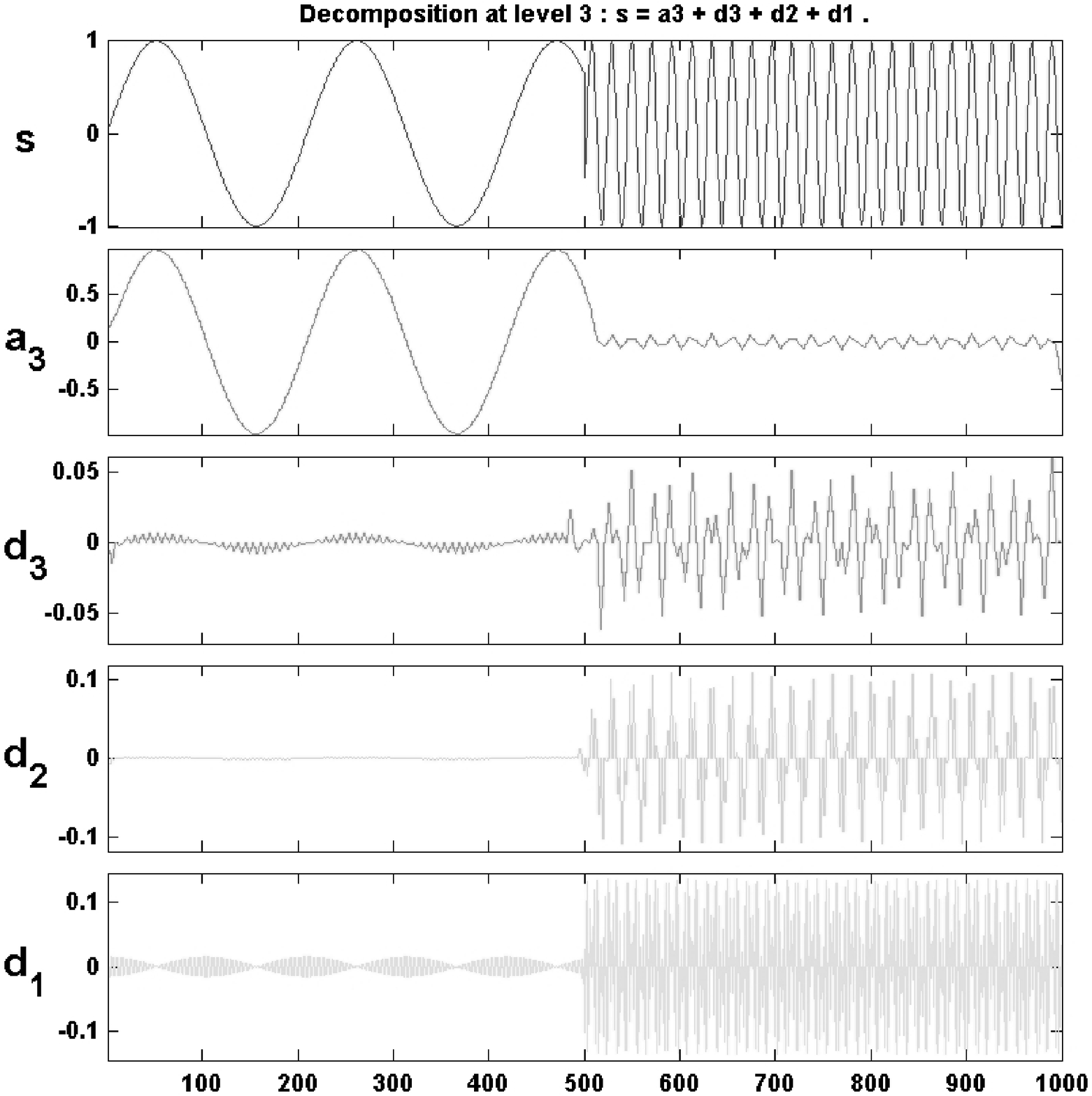,width=.45\linewidth}
} 
\subfigure[]{
\epsfig{file=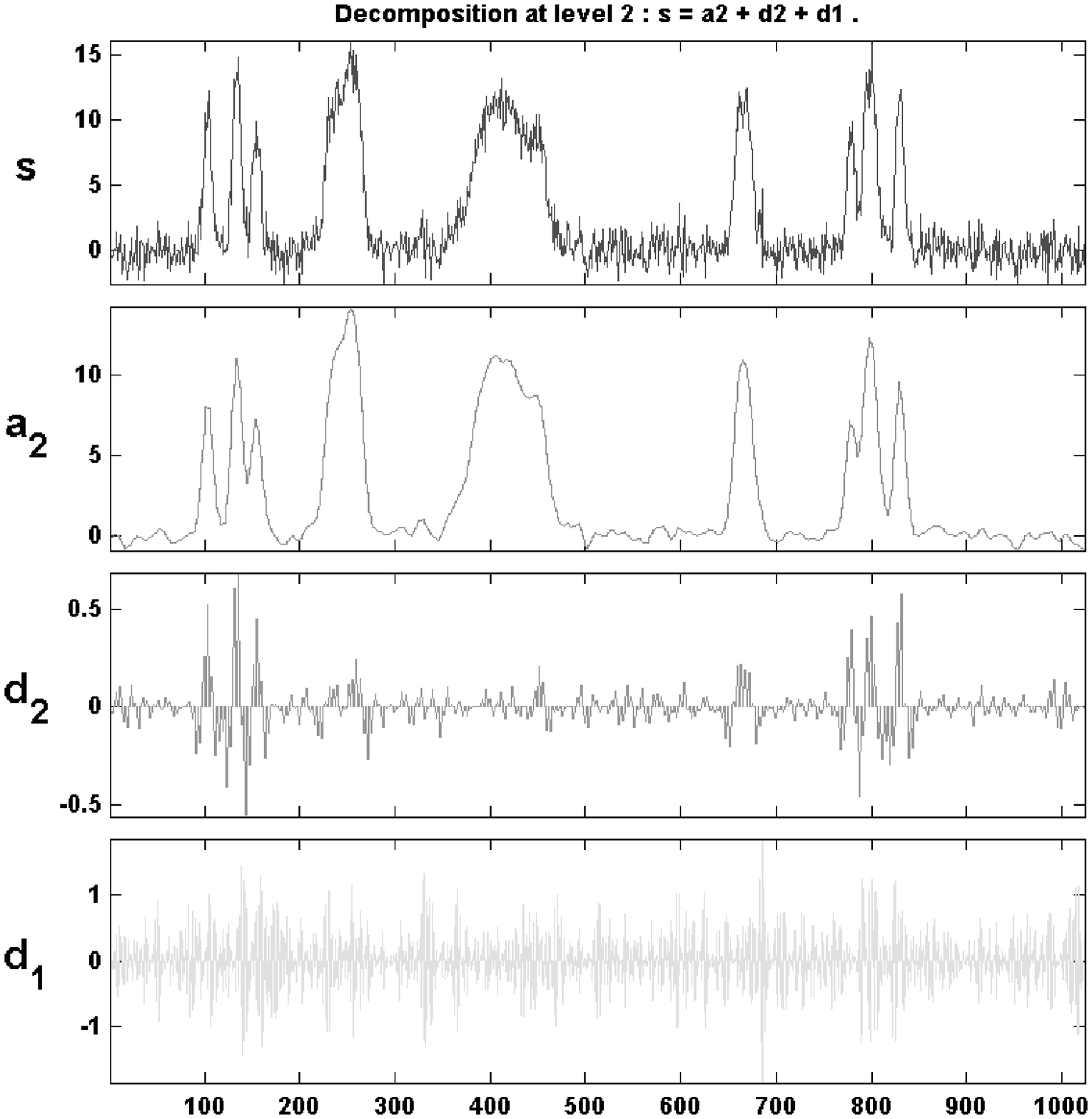,width=.45\linewidth}
}
\caption{(a) Analysis of a signal with a frequency breakdown (3-level decomposition),
(b) Denoising of {\tt noisbump} signal (2-level decomposition).
Both analysis were done with a wavelet generated by the Chebyshev polynomial of 2nd kind for
$m=3$.
These test signals are part of \matlab\ wavelet toolbox.}
\label{example:matlab}
\end{figure}

Proposed filters were implemented
in the \matlab\ Wavelet Toolbox~\cite{MisiMisi00}.
Standard sample signals were analyzed to illustrate
the behavior of the proposed wavelet and potential applications.

Figure~\ref{example:matlab} shows a 3-level decompositions 
of a standard frequency breakdown signal.
A standard noisy signal was also analyzed in a 2-level 
decomposition, 
illustrating potential uses 
of these wavelets in wave\-shrinkage~\cite{DonoJonh95}.

Impelled by a classical differential equation problem,
we introduced a new family of functions for signal analysis
via wavelet approach.
Based on the Chebyshev polynomials (type I and II) and
on the results derived in~\cite{Lira02},
we defined simple filter banks.

We showed that Chebyshev polynomials of 1st kind are not
naturally suitable wavelet construction via the cascade algorithm.
But on the other hand, we demonstrated that
the Chebyshev polynomials of 2nd kind are adequate 
for such an iterative process.
We also observed unexpected results, like the %
connection
between the magnitude of frequency response of the
filter based on Chebyshev polynomial of 2nd kind
and the well-know moving average filter.

The main properties of these filter banks were examined in detail.
In particular, a convergence proof for the iterative process with 
Chebyshev Type II filter banks was presented.
In Table~\ref{tab:prop}, we list a brief summary 
of the properties derived in this work.
Potential applications of
Chebyshev polynomials and wavelets are
particularly motivated by problems that deal with signal/pattern detection
or denoising.

\begin{table}
\centering
\caption{Summary of properties of Chebyshev filter banks.}
\label{tab:prop}

\begin{tabular}{lcc}
\toprule
Condition              & Type I         & Type II \\
\midrule
Symmetry               & Yes            & Yes \\
Perfect Reconstruction & Yes            & No  \\
Orthogonality          & Yes            & No  \\
Convergence\footnote{Filter iteration converges to wavelets.}            & No  & Yes \\
Compact Support        & ---            & Yes \\
\bottomrule
\end{tabular}
\end{table}

Finally we may call attention that 
the Chebyshev polynomials are in fact particular cases of
the more general Gegenbauer (ultraspherical) polynomials, 
which %
can be an attractive
tool for investigating new wavelet constructions.
Moreover, it is expected that Gegenbauer polynomials based wavelets
should exhibit a broader range of flexibility.

\section*{Acknowledgments}

This work was partially supported by CNPq and FACEPE.

\appendix

\section*{Proof of Proposition~\ref{proposition-ood-converges}}

We have that
$
h^{(2)}_m[n] = 
\frac{1}{m+1}
\big[
\underbrace{1\ \ 1\ \cdots\ 1\ \ 1
}_{\text{$m+1$ ones}}
\big]
$.
The rows of matrix $2\cdot\mathbf{H}\cdot\mathbf{H}^T$ have the following pattern
\begin{equation*}
\begin{split}
2
\frac{1}{m+1}\begin{bmatrix}1 & \cdots & 1 \end{bmatrix}
\ast
\frac{1}{m+1}\begin{bmatrix}1 & \cdots & 1 \end{bmatrix}
=
\frac{2}{(m+1)^2}\begin{bmatrix}1 & 2 & \cdots & m & m + 1 & m & \cdots & 2 & 1\end{bmatrix},
\end{split}
\end{equation*}
a triangular-shaped vector.
The matrix 
$\down 2\mathbf{H}\mathbf{H}^T$ is therefore described by:

\begin{equation*}
\begin{split}
\mathbf{T}_{2m-1}
= 
\frac{2}{(m+1)^2}  \cdot 
\left[
\begin{smallmatrix}
  &   &        &     &     &        &     &     &        &   &   \\ \medskip
2 & 1 & \cdots &  0  &  0  &  0     &  0  &  0  & \cdots & 0 & 0 \\ %
4 & 3 & \cdots &  0  &  0  &  0     &  0  &  0  &        & 0 & 0 \\ \medskip
  &   & \ddots &     &     & \ddots &     &     & \ddots &   &   \\ \medskip
6 & 7 & \cdots & m-1 & m-2 & m-3    & m-4 & m-5 & \cdots & 0 & 0 \\ \medskip
4 & 5 & \cdots & m+1 & m   & m-1    & m-2 & m-3 & \cdots & 1 & 0 \\ \medskip
2 & 3 & \cdots & m-1 & m   & m+1    & m   & m-1 & \cdots & 3 & 2 \\ \medskip
0 & 1 & \cdots & m-3 & m-2 & m-1    & m   & m+1 & \cdots & 5 & 4 \\ \medskip
0 & 0 & \cdots & m-5 & m-4 & m-3    & m-2 & m-1 & \cdots & 7 & 6 \\ \medskip
  &   & \ddots &     &     & \ddots &     &     & \ddots &   &   \\ %
0 & 0 & \cdots &  0  &  0  &  0     &  0  &  0  & \cdots & 3 & 4 \\ \medskip
0 & 0 & \cdots &  0  &  0  &  0     &  0  &  0  & \cdots & 1 & 2 \\ 
\end{smallmatrix}
\right].
\end{split}
\end{equation*}
One can check that such a specific matrix has the stochastic property: 
every column sums one. 
This can be done by separately analyzing even and odd columns, 
noting the fact that each column has even or odd elements only.
The sum of the columns of the even ($s_e$) and odd ($s_o$) elements
can be calculated by:
\begin{align}
s_e &= m+1 + 2\sum_{k=1}^{\frac{m-1}{2}} 2k
=
m+1 + 2 \frac{m-1}{2}\frac{m+1}{2} 
= \frac{(m+1)^2}{2},
\\
s_o &= 2\sum_{k=0}^{\frac{m-1}{2}} 2k+1 
= \frac{(m+1)^2}{2}.
\end{align}
Consequently, $\mathbf{T}_{2n-1}$ is a stochastic matrix.

The following theorem, derived from 
Perron-Frobenius Theorem~\cite[p.53]{Gant59},
is useful for showing that $\mathbf{T}_{2m-1}$ satisfies the conditions of
Theorem~\ref{CondE}.
\begin{theorem}[Eigenvalues of Irreducible Stochastic Matrix]
Let $\mathbf{M}$ be an irreducible Markov matrix. 
Then the number 1 is a simple eigenvalue of $\mathbf{M}$.
If $\mathbf{M}$ is aperiodic, then
$|\lambda|<1$ for all other eigenvalue $\lambda$ of $\mathbf{M}$.
\end{theorem}

It remains to show that $\mathbf{T}_{2m-1}$ is (a) irreducible and
(b) aperiodic.
The first condition is directly verified, 
because
$\mathbf{T}_{2m-1}$ is a band-like matrix with non null elements
within the band.
In Markov chain terminology, we can say that if all states can be reached from
each other, then $\mathbf{T}_{2m-1}$ is irreducible.
Moreover, the diagonal of matrix $\mathbf{T}_{2m-1}$ has all elements different
from zero, then all states have a self-loop. 
This guarantees that the periodicity
of the Markov matrix equals to 1 (aperiodicity).

\bibliographystyle{ieeetr}%

\bibliography{chebyshev}%

\begin{thebibliography}{10}

\bibitem{Arfk70}
G.~Arfken, {\em {Mathematical Methods for Physicists}}.
\newblock New York: Academic Press, 2nd~ed., 1970.

\bibitem{AbraSte68}
M.~Abramowitz and I.~Stegun, eds., {\em {Handbook of Mathematical Functions}}.
\newblock New York: Dover, 1968.

\bibitem{KilgPres94}
T.~Kilgore and J.~Prestin, ``{Polynomial Wavelets in the Interval},'' {\em
  Constructive Approximation}, vol.~12, pp.~95--110, 1996.
\newblock Springer-Verlag New York, Inc.

\bibitem{FiscPrest96}
B.~Fischer and J.~Prestin, ``{Wavelets Based on Orthogonal Polynomials},''
  1996.
\newblock Preprint.

\bibitem{Lira02}
M.~M.~S. Lira, H.~M. {de Oliveira}, and R.~J. {de Sobral Cintra},
  ``{Elliptic-Cylindrical Wavelets: The Mathieu Wavelets},'' {\em IEEE Signal
  Processing Letters}, 2003.
\newblock To appear.

\bibitem{Lira03}
M.~M.~S. Lira, H.~M. {de Oliveira}, and R.~M. {Campello de Souza}, ``{New
  Orthogonal Compact Support Wavelet Derived from Legendre Polynomials:
  Spherical Harmonic Wavelets}.'' To be submitted.

\bibitem{Mall89}
S.~Mallat, ``{A Theory for Multiresolution Signal Decomposition: The Wavelet
  Representation},'' {\em IEEE Transactions on Pattern Analysis and Machine
  Intelligence}, vol.~11, pp.~674--693, July 1989.

\bibitem{Oppe99}
A.~V. Oppenheim and R.~W. Schafer, {\em {Discrete-time Signal Processing}}.
\newblock New Jersey: Prentice-Hall, 1999.

\bibitem{MisiMisi00}
M.~Misiti, Y.~Misiti, G.~Oppenheim, and J.-M. Poggi, {\em {Wavelet Toolbox
  User's Guide}}.
\newblock New York: The MathWorks, Inc., 2nd~ed., 2000.

\bibitem{VettKova95}
M.~Vetterli and J.~Kova\v{c}evi\'c, {\em {Wavelets and Subband Coding}}.
\newblock New Jersey: Prentice-Hall, 1995.

\bibitem{StraNguy96}
G.~Strang and T.~Nguyen, {\em {Wavelets and Filter Banks}}.
\newblock Wellesley: Wellesley-Cambridge Press, 1996.

\bibitem{Vett01}
M.~Vetterli, ``{Wavelets, Approximation, and Compression},'' {\em IEEE Signal
  Processing Magazine}, pp.~59--73, Sept. 2001.

\bibitem{SmitBarn86}
M.~J.~T. Smith and T.~P. {Barnwell, III}, ``{Exact Reconstruction Techniques
  for Tree-Structured Subband Coders},'' {\em IEEE Transactions on Acoustics,
  Speech, and Signal Processing}, vol.~34, pp.~434--441, June 1986.

\bibitem{Lawt91}
W.~M. Lawton, ``{Necessary and Sufficient Conditions for Constructing
  Orthonormal Wavelet Bases},'' {\em Journal of Mathematical Physics}, vol.~32,
  pp.~57--61, Jan. 1991.

\bibitem{DonoJonh95}
D.~L. Donoho and I.~M. Johnstone, ``{Adapting to Unknown Smoothness via Wavelet
  Shrinkage},'' {\em Journal of the American Statistical Association}, vol.~90,
  no.~432, pp.~1200--1224, 1995.

\bibitem{Gant59}
F.~R. Gantmacher, {\em {The Theory of Matrices}}, vol.~2.
\newblock New York: Chelsea, 1959.

\end{thebibliography}

\end{document}